\documentclass{article}
\usepackage{arxiv}

\usepackage[utf8]{inputenc} 
\usepackage[T1]{fontenc}    
\usepackage{hyperref}       
\usepackage{url}            
\usepackage{booktabs}       
\usepackage{amsfonts}       
\usepackage{nicefrac}       
\usepackage{microtype}      
\usepackage{lipsum}		
\usepackage{graphicx}
\usepackage{natbib}
\usepackage{doi}
\usepackage{amsmath,amssymb,amsfonts}%
\usepackage{amsthm}%
\usepackage{mathrsfs}%
\usepackage{bbm}%
\usepackage{textcomp}%
\usepackage{manyfoot}%
\usepackage{bm}%
\usepackage{threeparttable}%
\usepackage{color}%

\DeclareMathOperator*{\argmax}{arg\,max}

\title{A stochastic method to estimate a zero-inflated two-part mixed model for human microbiome data}

\date{} 					

\author{\href{https://orcid.org/0009-0005-5811-7062}{\includegraphics[scale=0.06]{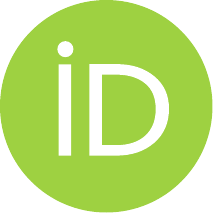}\hspace{1mm}
    John Barrera}
    \thanks{Corresponding author. E-mail: \texttt{john.barrera@postgrado.uv.cl}} \\
	  Instituto de Ingeniería Matemática \\
    Facultad de Ingeniería \\
    Universidad de Valparaíso \\
    Valparaíso, Chile\\
	\And
	\href{https://orcid.org/0000-0001-6332-6137}{\includegraphics[scale=0.06]{pix/orcid.pdf}\hspace{1mm}
    Cristian Meza} \\
	CIMFAV \\
    Universidad de Valparaíso \\
    Valparaíso, Chile\\
	\And
    \href{https://orcid.org/0000-0002-2398-3156}{\includegraphics[scale=0.06]{pix/orcid.pdf}\hspace{1mm}
    Ana Arribas-Gil} \\
	Departamento de Estadística \\ 
    Universidad Carlos III de Madrid \\
    Getafe, Spain\\
}

\renewcommand{\headeright}{}
\renewcommand{\undertitle}{}
\renewcommand{\shorttitle}{Stochastic estimation of a zero-inflated mixed model for microbiome data}

\hypersetup{
pdftitle={A stochastic method to estimate a zero-inflated two-part mixed model for human microbiome data},
pdfauthor={J. Barrera et al.}
}

\begin{document}
\maketitle

\begin{abstract}
	Human microbiome studies based on genetic sequencing techniques produce longitudinal data of the relative abundances of microbial taxa over time, allowing to analyse, through mixed-effects modeling, how microbial communities evolve in response to clinical interventions, environmental changes, or disease progression. In particular, the Zero-Inflated Beta Regression (ZIBR) models jointly and over time the presence and abundance of each microbe taxon, considering the bounded nature of the data, its skewness, and the over-abundance of zeros. However, as for other complex random effects models, maximum likelihood estimation suffers from the intractability of likelihood integrals. Available estimation methods rely on log-likelihood approximation, prone to limitations such as biased estimates or unstable convergence. In this work we develop an alternative maximum likelihood estimation approach for the ZIBR model, via the Stochastic Approximation Expectation Maximization (SAEM) algorithm. The proposed methodology can accommodate unbalanced data, not always possible in existing approaches. We also provide estimations of standard errors and log-likelihood of the fitted model. The performance of the algorithm is established through simulation, and its use is demonstrated on two microbiome studies, showing its ability to detect changes in presence and abundance of bacterial taxa over time and in response to treatment.
\end{abstract}


\section{Introduction}\label{sec1}

The human microbiome, a complex community of microorganisms, plays a crucial role in the body's functions and health. It influences metabolic pathways, drug metabolism, and contributes to the bioconversion of nutrients, detoxification, and protection against pathogens \citep{biblio1}. In particular, the gut microbiota has been shown to interact with the host immune system, influencing the development of some diseases \citep{biblio4}. The microbiome significantly impacts the human genome, with genotypes influencing its composition and activity, while the microbiome alters the expression of genetic risk for chronic inflammatory and immune conditions \citep{biblio5}.

Accurate analysis of microbiome data requires reliable collection and processing frameworks. High-throughput sequencing technologies, such as the 16S ribosomal RNA (rRNA) and shotgun sequencing, underlie large initiatives such as the Human Microbiome Project \citep{biblio6}. These methods generate sequence counts whose totals depend on instrument capacity, motivating normalization and the resulting compositional data structure \citep{biblio7}. This constant-sum constraint is key for comparing taxa distributions across samples, as in differential abundance analysis. However, when studying the marginal behavior of individual taxa over time or under interventions, it is common to model their proportions as bounded outcomes in $(0,1)$. In the microbiome literature, such longitudinal proportion data are often referred to as compositional even when analyzed taxon-wise, and we adopt that terminology throughout this paper.

Regarding data analysis, \cite{biblio8} compiled some recent models for longitudinal microbiome data based on counts or relative abundances. Among the latter, the Zero-Inflated Beta Regression (ZIBR) model \citep{biblio9}, built on the work of \cite{biblio55}, provides a mixed-effects two-part framework to jointly model presence/absence allowing the inclusion of clinical covariates, both to explain the presence or absence and, conditional on presence, relative abundance. It naturally accommodates boundedness, skewness, and zero inflation. Since its introduction, ZIBR has been used successfully in several studies \citep{biblio11,biblio10}, capturing within-subject correlation and enabling inference on covariate effects.

As for other complex mixed-effects models, ZIBR estimation via maximum likelihood (ML) relies on Gauss–Hermite quadrature and numerical optimization. Yet such approximations may introduce bias, and alternative methods often outperform ML in complex generalized mixed models \citep{biblio13}. Moreover, the proposed ML implementation is suited only for balanced data, whereas clinical studies frequently yield unbalanced or missing observations, potentially leading to biased conclusions \citep{biblio15,biblio14}. The ZIBR model can also be thought of as a particular case of the Generalized Additive Model for Location, Scale and Shape (GAMLSS, \citealp{biblio52}; \citealp{GAMLSS}), which handles missing data but still relies on penalized local-likelihood approximations that complicate likelihood ratio testing in complex GLMMs (see \cite{biblio56}).

All things considered, the versatile features of the ZIBR model make it a promising choice for the analysis of proportion microbiome data, despite possible drawbacks in existing estimation strategies. Therefore, for the final purpose of precisely identifying those taxa responsive to disease onset, changes in environmental conditions or specific interventions, any possible improvement in the estimation method which is able to provide more accurate estimates will amount to significant progress in the understanding of human microbiome and its relation to human health. Following this aim, in this work we propose a new estimation framework for the ZIBR model on longitudinal proportion data based on the Stochastic Approximation EM (SAEM) algorithm \citep{biblio20}, an exact ML strategy for missing-data models where classical EM (\citealp{biblio18}; \citealp{McLachlanKrishnan2008}) is not directly applicable due to intractable conditional expectations, precisely the case of ZIBR. The SAEM algorithm preserves the convergence and monotonicity properties of EM, has shown strong performance in complex mixed models \citep{MOD:2012,biblio21}, provides inference and hypothesis-testing tools \citep{biblio22}, and can be enhanced via MCMC techniques \citep{biblio23} or extended to Restricted Maximum Likelihood (REML) estimation \citep{biblio25}.

In this article we extend the algorithm to distributions not belonging to the exponential family and derive the explicit expressions at all its steps, for both parameter estimation and log-likelihood approximation, once the ML estimators have been obtained. We also obtain approximations of the standard errors of the estimators, by means of the stochastic approximation of the Fisher information matrix. This allows us to provide a comprehensive estimation approach that avoids downsides related to likelihood approximations, is able to incorporate unbalanced data, and facilitates the inference pipeline, from modeling to covariate effects testing, under the same framework.

Notice that, although the motivation for this proposal arises from the analysis of microbiome data, the potential applications of the ZIBR model extend far beyond this context. For example, in \cite{biblio62} it is used to study the technical efficiency of hospitals in Spain based on their ownership structure, while in \cite{biblio63} it is used to analyze data on plant family coverage in South Africa, adding spatial components. These are just a few examples of its possible applications.

The structure of the document is as follows: in Section \ref{sec2}, we introduce the ZIBR model and develop the SAEM based inference method to be used in our work. In Section \ref{sec3} we present simulation studies on synthetic data generated under different settings, comparing the results obtained with our approach and those given by estimation based on likelihood approximation or penalization, and in Section \ref{sec4} we assess the behaviour of the proposed routine on a dataset coming from clinical microbiome studies. Finally, Section \ref{sec5} closes the article with the main conclusions, a discussion of the results, and possible limitations and future developments.

\section{Models and methods}\label{sec2}

In this section we describe the ZIBR model for longitudinal proportion data \citep{biblio9}, we revise the foundations of the SAEM algorithm for parameter estimation by maximum likelihood as well as log-likelihood estimation through Importance Sampling, and present its extension to the ZIBR model.

\subsection{The ZIBR model for longitudinal proportion data}\label{sec21}

The ZIBR model describes the presence and abundance of a single bacterial taxon on different individuals over time, and can be subsequently applied to different bacteria. Let $y_{it}$ be the relative abundance of a bacterial taxon in the individual $i$ at time $t$, $1\leq i\leq N$, $1\leq t\leq T_i$. The model assumes that $y_{it}$ follows the distribution:
\begin{equation}\label{zibr1}
    y_{it}\sim
    \begin{cases}
    0&\mbox{with prob. } 1-p_{it},\\
    Beta(u_{it}\phi,(1-u_{it})\phi)&\mbox{with prob. } p_{it}\\
    \end{cases}    
\end{equation}
with $\phi>0$ and $0<u_{it},p_{it}<1$. These two last components are characterized by
\begin{equation}\label{zibr2}
    \log{\left(\frac{p_{it}}{1-p_{it}}\right)}=a_i+\bm{X}_{it}^{\top}\bm{\alpha},\hspace*{1cm}\log{\left(\frac{u_{it}}{1-u_{it}}\right)}=b_i+\bm{Z}_{it}^{\top}\bm{\beta},    
\end{equation}
where $a_i$ and $b_i$ are individual specific intercepts, $\bm{\alpha}$ and $\bm{\beta}$ are vectors of regression coefficients and $\bm{X}_{it}$ and $\bm{Z}_{it}$ are covariates for each individual and time point. We further consider that each one of the random intercepts follows a normal distribution, independently from each other:
$a_i\sim N(a,\sigma^2_1),b_i\sim N(b,\sigma^2_2).$
From Equations \ref{zibr1} and \ref{zibr2}, it can be seen that the ZIBR model explicitly includes a component that is responsible for the presence of zeros in the data. It is also clear that conveniently defined covariates $\bm{X}_{it}$ and $\bm{Z}_{it}$ can influence both the probability of presence or absence of a bacterial taxon (through the logistic regression that defines $p_{it}$) and the magnitude of its relative abundance (through the $u_{it}$ component in the proposed Beta distribution). Furthermore, the inclusion of a random intercept allows modeling correlations in observations from the same individual. Even though it is easy to expand the definition to consider random slopes, in practice it is enough to consider just a random intercept \citep{biblio26}.

The model parameter $\bm{\theta}=(\phi,a,b,\bm{\alpha},\bm{\beta},\sigma^2_1,\sigma^2_2)$ can be estimated by maximum likelihood. From Equations \ref{zibr1} and \ref{zibr2}, the likelihood function for data $\bm{y}=\left(y_{it}, 1\leq i\leq N, 1\leq t\leq T_i\right)$ is
\begin{equation}\label{zibrlik}
      L(\bm{\theta};\bm{y})=\prod_{i=1}^N \int_{\mathbb{R}}\int_{\mathbb{R}} \prod_{t=1}^{T_i} (1-p_{it})^{\mathbbm{1}_{\left\{y_{it}=0\right\}}}\left[p_{it}f(y_{it};u_{it},\phi)\right]^{\mathbbm{1}_{\left\{y_{it}>0\right\}}}g\left(a_i,b_i|a,\sigma^2_1,b,\sigma^2_2\right) da_i db_i      
\end{equation}    
where $f(y_{it};u_{it},\phi)$ is the Beta density function with parameters $u_{it}$ and $\phi$ on $y_{it}$: 
$$f(y_{it};u_{it},\phi)=\frac{\Gamma(\phi)}{\Gamma(u_{it}\phi)\Gamma((1-u_{it})\phi)} y_{it}^{u_{it}\phi-1}(1-y_{it})^{(1-u_{it})\phi-1},$$
and $g$ is the product of the two univariate normal density functions of random effects $a_i$ and $b_i$. 

Given the impossibility of analytical calculation of the integral shown in Equation \ref{zibrlik}, an approximation can be achieved by means of the Gauss-Hermite quadrature (GHQ). With this approximation, and through numerical optimization, the maximum likelihood estimators for $\theta$ can be found as proposed by \cite{biblio9}. Hypothesis tests for the significance of covariates can also be conducted, in particular the Likelihood Ratio Test (LRT). The implementation of this approach is available in the \texttt{ZIBR} package \citep{biblio27} developed for the R software. In addition to this alternative, the \texttt{gamlss} package \citep{biblio52} can also be used, which, in a similar manner to the aforementioned one, is based on a penalized quasilikelihood approximation and its optimization based on numerical algorithms \citep{biblio53}. A notable advantage of this implementation, however, is its capacity to handle unbalanced data, which renders it a suitable option for a comparative analysis of the results obtained in this study. 

\subsection{The SAEM algorithm for mixed effects models}\label{sec22}

The Stochastic Approximation Expectation-Maximization (SAEM) algorithm \citep{biblio20} is a powerful tool for estimating population parameters in complex mixed effect models. This algorithm is applicable for the iterative computation of ML estimates in a wide variety of incomplete data statistical problems in which the Expectation step of the EM algorithm is not explicit; in particular in mixed effects models, where the individual random effects are treated as non-observed data. Let $\bm{y}=\left(y_{it}, 1\leq i\leq N, 1\leq t\leq T_i\right)$ and $\boldsymbol{\varphi}=\left(\varphi_i, 1\leq i\leq N\right)$ denote the observed and non-observed data, respectively, so the complete data of the model are $(\bm{y},\boldsymbol{\varphi})$. In this case, the SAEM algorithm consists of replacing the usual E-step of EM with a stochastic approximation procedure. Given an initial point $\bm{\theta}^{(0)}$, iteration $q$ of the algorithm writes:
\begin{itemize}
    \item \textbf{Simulation (S) step:} Draw a realization $\boldsymbol{\varphi}^{(q)}$ from the conditional distribution $p\left(\cdot|\;\bm{y};\bm\theta^{(q-1)}\right)$.
    \item \textbf{Stochastic Approximation (SA) step:} Update $s_{q}(\bm\theta)$, the approximation of the conditional expectation $\mathbb{E}\left[\log p\left(\bm{y},\bm{\varphi}^{(q)};\bm\theta\right)|\bm{y},\bm\theta^{(q-1)}\right]$:
    $$s_q(\bm\theta)=s_{q-1}(\bm\theta) + \gamma_q \left(\log p\left(\bm{y}, \bm{\varphi}^{(q)}; \bm\theta\right)- s_{q-1}(\bm\theta)\right)$$
\noindent where $\{\gamma_q\}_{q\in\mathbb{N}}$ is a decreasing sequence of stepsizes with $\gamma_1=1$. 
    \item \textbf{Maximization (M) step:} Update $\bm\theta^{(q)}$ according to $\bm\theta^{(q)}= \argmax_{\bm\theta} s_q(\bm\theta).$
\end{itemize}
There are some important remarks on the the working details of the SAEM algorithm. In the case of complex mixed effects models, such as ZIBR, the conditional distribution of the non-observed data $p\left(\cdot|\;\bm{y};\bm\theta^{(q-1)}\right)$ cannot be computed in closed form and simulation from it cannot be carried out directly \citep{biblio23}. However, a MCMC approach can be used in the Simulation step of the SAEM algorithm described above, consisting in applying the Metropolis-Hastings algorithm \citep{biblio31} with different proposal kernels, in order to approximate $p\left(\cdot|\;\bm{y};\bm\theta^{(q-1)}\right)$ with a Markov chain with defined transition probabilities. 

Also, convergence can be improved by generating more than one Markov chain or realization at simulation and by applying a Monte Carlo scheme. That is, at the Simulation step $m$ realizations $\boldsymbol{\varphi}^{(q,l)}\sim p\left(\cdot|\;\bm{y};\bm\theta^{(q-1)}\right)$, $1\leq l\leq m$, are drawn, and in the SA step the approximation of the conditional expectation is updated as $$s_q(\bm\theta)=s_{q-1}(\bm\theta) + \gamma_q \left(\frac{1}{m}\sum_{l=1}^m \log p\left(\bm{y}, \bm{\varphi}^{(q,l)}; \bm\theta\right)- s_{q-1}(\bm\theta)\right).$$

Let $\langle a, b\rangle$ denotes the standard inner product between vectors $a$ and $b$. If the complete-data model belongs to the exponential family, that is, if 
$$\log p(\bm{y},\boldsymbol{\varphi};\bm\theta)=-\Psi(\bm\theta)+\langle
S(\bm{y},\boldsymbol{\varphi}),\xi(\bm\theta)\rangle$$
where $S(\bm{y},\boldsymbol{\varphi})$ represents a sufficient statistic of the data, then, the SA step reduces to:
\begin{equation}\label{SA_step_exp} 
F_q=F_{q-1} + \gamma_q \left( \frac{1}{m}\sum_{l=1}^m S(\bm{y},\boldsymbol{\varphi}^{(q,l)})- F_{q-1}\right)
\end{equation}
given that $s_q(\bm\theta)=-\Psi(\bm\theta)+\langle F_q(\bm{y},\boldsymbol{\varphi}),\xi(\bm\theta)\rangle$; that is, the actualization is made only on the sufficient statistic. This scheme can be applied even to models outside the exponential family, provided that a part of the model belongs to this family. However, we cannot speak of updating a sufficient statistic, but rather of a data summary function. Under general circumstances \citep{biblio20,biblio23}, the convergence of the parameter sequence $\bm\theta^{(q)}$ toward a (local) maximum of the likelihood $\hat{\bm\theta}$ is guaranteed, regardless of the starting point $\bm\theta^{(0)}$ \citep{biblio50}.

The sequence of stepsizes $\{\gamma_q\}_{q\in\mathbb{N}}$ is usually set to 1 during the first iterations to avoid getting stuck in local maxima. In this way the first iterations are identical to those of the Monte Carlo EM (MCEM) algorithm \citep{biblio51}, which is known for its slow convergence rate. To avoid this scenario, in later iterations of SAEM $\gamma_q$ decreases to zero to force convergence with fewer iterations. Details of application of the SAEM algorithm to complex mixed-effects models can be found in \cite{MOD:2012, ABMR:2014, biblio21} and \cite{DLMN:2024}.

\subsection{The SAEM algorithm for ZIBR parameter estimation}\label{sec23}

As we have seen before, a mixed model can be considered as an unobserved data problem and therefore be addressed using the SAEM algorithm. Let us consider $\bm\varphi_i=(a_i,b_i)$, $1\leq i \leq N$, the non-observed data. By the definition of the ZIBR model $\bm\varphi_i$ follows the multivariate normal distribution $\bm\varphi_i\sim N(\boldsymbol{\mu},\mathbf{G})$ with $\boldsymbol{\mu}=(a,b)$ and $\mathbf{G}=diag(\sigma_1^2,\sigma_2^2)$. With the usual notation $\bm{y}=(y_{it}:\ 1\leq i\leq N,1\leq t\leq T_i)$ and $\boldsymbol{\varphi}=(\bm\varphi_{i}:\ 1\leq i\leq N)$, the complete-data likelihood writes:
\begin{equation}\label{decomp}
    \begin{split}
    p(\bm{y},\boldsymbol{\varphi};\theta)= &p(\bm{y}|\boldsymbol{\varphi};\alpha,\beta,\phi)p(\boldsymbol{\varphi}|\boldsymbol{\mu},\mathbf{G})\\
    \propto& |\mathbf{G}|^{-\frac{N}{2}} \prod_{i} \exp{\left(-\frac{(\bm\varphi_i-\boldsymbol{\mu})^T\mathbf{G^{-1}}(\bm\varphi_i-\boldsymbol{\mu})}{2}\right)}\\ 
    &\times\prod_{i,t} (1-p_{it})^{\mathbbm{1}_{\left\{y_{it}=0\right\}}}p_{it}^{\mathbbm{1}_{\left\{y_{it}>0\right\}}}f(y_{it};u_{it},\phi)^{\mathbbm{1}_{\left\{y_{it}>0\right\}}}    .
\end{split}
\end{equation}
Like most zero-inflated models, the ZIBR model cannot be considered part of the exponential family \citep{biblio48}. However, the decomposition presented in Equation \ref{decomp} allows us to propose a simplified structure for the SAEM algorithm (Equation \ref{SA_step_exp}). For the multivariate normal part corresponding to the random effects, the actualization in the SA step is done on the respective sufficient statistics. For the mixture distribution corresponding to the observed data, $\bm{y}|\boldsymbol{\varphi};\bm\alpha,\bm\beta,\phi$, maximization of the conditional log-likelihood is followed by estimates updates, as suggested for non-exponential family models \citep{saemix_userguide}. Then, the Maximum Likelihood iterative estimation algorithm for the parameters of the ZIBR model writes, for a given starting point $\bm\theta^{(0)}$, the summary data functions $F_1^{(0)}(\bm{y},\boldsymbol{\varphi})=\sum_i {\bm\varphi_i^{(0)}}$ and $F_2^{(0)}(\bm{y},\boldsymbol{\varphi})=\sum_i {\bm\varphi_i^{(0)}{\bm\varphi_i^{(0)}}^T}$ and at iteration $q$, as:
\begin{enumerate}
    \item \textbf{Simulation step:} draw $\bm\varphi^{(q)}_i,i=1,\cdots,N$ from the distribution $p(\cdot|\bm{y};\bm\theta^{(q-1)})$.
    \item \textbf{Stochastic Approximation step:} update the summary data functions with the scheme:
    \begin{equation}\label{update}
    \begin{split}
    F_1^{(q)}(\bm{y},\boldsymbol{\varphi})&=F_1^{(q-1)}(\bm{y},\boldsymbol{\varphi})+\gamma_q \left(\sum_i {\bm\varphi_i^{(q)}}-F_1^{(q-1)}(\bm{y},\boldsymbol{\varphi})\right)\\
    F_2^{(q)}(\bm{y},\boldsymbol{\varphi})&=F_2^{(q-1)}(\bm{y},\boldsymbol{\varphi})+\gamma_q \left(\sum_i {\bm\varphi_i^{(q)}{\bm\varphi_i^{(q)}}^T}-F_2^{(q-1)}(\bm{y},\boldsymbol{\varphi})\right).
    \end{split}
    \end{equation}
    where $\{\gamma_q\}_{q\in\mathbb{N}}$ is a decreasing sequence of stepsizes with $\gamma_1=1$. 
    \item \textbf{Maximization step:} update the parameters of the model with
    \begin{equation}\label{maxim}
    \begin{split}
    \boldsymbol{\mu}^{(q)}&=\frac{F_1^{(q)}(\bm{y},\boldsymbol{\varphi})}{N}\\
    \mathbf{G}^{(q)}&=\frac{F_2^{(q)}(\bm{y},\boldsymbol{\varphi})}{N}-\frac{F_1^{(q)}(\bm{y},\boldsymbol{\varphi}){F_1^{(q)}(\bm{y},\boldsymbol{\varphi})}^T}{N^2}.
    \end{split}
    \end{equation}
    \end{enumerate}
    Given the form of the model definition in the Beta part, steps 2 and 3 are modified by first calculating
    \begin{eqnarray}\label{update1}
        \left(\tilde{\bm\beta}^{(q)},\tilde{\phi}^{(q)}\right)=&\\
        \argmax_{\bm\beta,\phi}\sum_{i,t}&\left[\mathbbm{1}_{\left\{y_{it}>0\right\}}\left(\log{\frac{\Gamma(\phi)}{\Gamma\left(u_{it}^{(q)}\phi\right)\Gamma\left(\left(1-u_{it}^{(q)}\right)\phi\right)}}+u_{it}^{(q)}\phi\log{y_{it}}+\phi\left(1-u_{it}^{(q)}\right)\log{(1-y_{it})}\right) \right]\notag
    \end{eqnarray}
    and
    \begin{equation}\label{update2}
        \tilde{\bm\alpha}^{(q)}=\argmax_{\bm\alpha}\sum_{i,t} \left[\mathbbm{1}_{\left\{y_{it}>0\right\}}\log{\left(p_{it}^{(q)}\right)}+\mathbbm{1}_{\left\{y_{it}=0\right\}}\log{\left(1-p_{it}^{(q)}\right)}\right]    
    \end{equation}
    where $u_{it}^{(q)}=u_{it}^{(q)}(b_i,\bm\beta)$ and $p_{it}^{(q)}=p_{it}^{(q)}(a_i,\bm\alpha)$ are calculated using $\varphi^{(q)}_i$ and Equation \ref{zibr2}. Maximization in (\ref{update1}) and (\ref{update2}) is achieved numerically. Finally, the values are updated by doing
    \begin{equation}\label{maxim1}
    \begin{split}
        \phi^{(q)}&=\phi^{(q-1)}+\gamma_{q}\left(\tilde{\phi}^{(q)}-\phi^{(q-1)}\right)\\
        \bm\alpha^{(q)}&=\bm\alpha^{(q-1)}+\gamma_{q}\left(\tilde{\bm\alpha}^{(q)}-\bm\alpha^{(q-1)}\right);\quad
        \bm\beta^{(q)}=\bm\beta^{(q-1)}+\gamma_{q}\left(\tilde{\bm\beta}^{(q)}-\bm\beta^{(q-1)}\right)
    \end{split}
    \end{equation}
Let us discuss the details of this implementation. As mentioned in \ref{sec22}, the choice of the starting point $\bm\theta^{(0)}$ for SAEM does not affect its convergence; however, it is recommended to use values obtained in previous studies or with other estimation methods. Following the example of the existing implementation of the \texttt{saemix} package \citep{saemix}, we will use $\gamma_q=1$ if $q\leq K_1$ , and  $\gamma_q=\frac{1}{q-K_1}$ if $K_1<q\leq K_1+K_2$, where $K_1+K_2$ is the total number of iterations. 

We also discussed in section \ref{sec22} that it is possible to improve the performance of the algorithm by taking multiple sequences or Markov chains in the Simulation step, and using Monte Carlo in Equations \ref{maxim} and \ref{maxim1}. Furthermore, during the SA step, we obtain sequences that allow to estimate $\mathbb{E}\left(\bm\varphi_i | y_i ; \hat{\bm\theta}\right)$ and $\mbox{Var}\left(\bm\varphi_i|y_i; \hat{\bm\theta}\right)$ to be calculated, values necessary to approximate the log-likelihood through Importance Sampling, with which the Likelihood Ratio Test (LRT) can be computed, as presented in the following subsection. 

\subsubsection{Approximation of the log-likelihood using Importance Sampling}\label{sec231}

The log-likelihood of the observed data cannot be computed in closed form for complex mixed effects models, but its estimation is required to perform the LRT and to compute information criteria for a given model. One approximation method is given by the application of the Importance Sampling algorithm \citep{Kloek:1978}. Let $\mathcal{L}\mathcal{L}_y(\hat{\bm\theta})$ be the log-likelihood of the model at the vector of population parameter estimates, that is $\mathcal{LL}_y(\hat{\bm\theta})=\log p(\bm{y};\hat{\bm\theta})$ where $p(\bm{y};\hat{\bm\theta})=L(\hat{\bm\theta}; \bm{y})$ is the joint probability distribution function of the observed data given $\hat{\bm\theta}$. Notice that $\mathcal{LL}_y(\hat{\bm\theta})=\log p(\bm{y};\hat{\bm\theta})=\sum_{i=1}^N \log p(y_i;\hat{\bm\theta})$
and, for some \textit{proposal distribution} $\tilde{p}_{\bm\varphi_i}$ absolutely continuous with respect to $p_{\bm\varphi_i}$, we have
$$p(y_i;\hat{\bm\theta})=\int p(y_i,\bm\varphi_i;\hat{\bm\theta})d\bm\varphi_i
=\int p(y_i|\bm\varphi_i;\hat{\bm\theta})\frac{p(\bm\varphi_i;\hat{\bm\theta})}{\tilde{p}(\bm\varphi_i;\hat{\bm\theta})} \tilde{p}(\bm\varphi_i;\hat{\bm\theta}) d\bm\varphi_i
=\mathbb{E}_{\tilde{p}} \left[p(y_i|\bm\varphi_i;\hat{\bm\theta})\frac{p(\bm\varphi_i;\hat{\bm\theta})}{\tilde{p}(\bm\varphi_i;\hat{\bm\theta})}\right].$$
That is, $p(y_i;\hat{\bm\theta})$ can be expressed as an expectation which can be approximated by:
\begin{enumerate}
    \item Obtain a random sample $\bm\varphi_i^{(1)},\cdots,\bm\varphi_i^{(K)}$ from the proposal distribution $\tilde{p}_{\bm\varphi_i}$;
    \item Compute the empirical mean $\hat{p}_{(i,K)}=\frac{1}{K}\sum_{k=1}^K p(y_i|\bm\varphi_i^{(k)};\hat{\bm\theta})\frac{p(\bm\varphi_i^{(k)};\hat{\bm\theta})}{\tilde{p}(\bm\varphi_i^{(k)};\hat{\bm\theta})}$
\end{enumerate}
An optimal proposal distribution would be the conditional distribution $p_{\bm\varphi_i|y_i}$ since in that case the estimator of the expectation has zero variance. But since the closed form expression of the distribution is not available, we choose a proposal \textit{close} to this optimal distribution, based on the empirically estimated conditional mean and variance, $\mu_i=\hat{\mathbb{E}}\left[\bm\varphi_i|y_i;\hat{\bm\theta}\right]$ and $\sigma^2_{i}=\hat{\mbox{Var}}\left[\bm\varphi_i|y_i;\hat{\bm\theta}\right]$, of the simulated random effects during the simulation step of the SAEM algorithm. Then, the proposed candidate $\bm\varphi_i^{(k)}$, with $k=1,\dots,K$, is drawn with a noncentral Student $t$-distribution $\bm\varphi_i^{(k)}=\mu_i + \sigma_{i} \times T_{i,k}$,
with $T_{i,k}\sim t_{\nu}$ i.i.d., where $t_{\nu}$ denotes a Student t-distribution with $\nu$ degrees of freedom. In this work, unless otherwise mentioned, the parameters for calculating the log-likelihood will be $\nu=5$ and $K=500$.

\subsubsection{Stochastic approximation of the standard errors}\label{sec232}

In addition to providing estimates of the parameters of a model, it is desirable that the estimation method is capable of also estimating its standard errors, with the objective of constructing confidence intervals or performing statistical tests on individual estimators, such as the Wald test. In the case of maximum likelihood estimation, these errors can asymptotically be calculated based on the Fisher information matrix of the model, which for complex models cannot be computed in a closed form. Based on the Louis's missing information principle \citep{biblio41} it is possible to compute an estimation of the Fisher information matrix. According to this principle, we have the identity:
$$\partial_{\theta}^2\log p(\bm{y};\bm\theta)=\mathbb{E}\left(\partial_{\bm\theta}^2\log p(\bm{y},\bm{\varphi};\bm\theta)|\bm{y};\bm\theta\right)+\mbox{Cov}\left(\partial_{\bm\theta}\log p(\bm{y},\bm{\varphi};\bm\theta)|\bm{y};\bm\theta\right)$$
where
\begin{align*}
  \mbox{Cov}\left(\partial_{\bm\theta}\log p(\bm{y},\bm{\varphi};\bm\theta)|\bm{y};\bm\theta\right) =&\mathbb{E}\left(\partial_{\bm\theta}\log p(\bm{y},\bm{\varphi};\bm\theta)\partial_{\bm\theta}\log p(\bm{y},\bm{\varphi};\bm\theta)^{\top}|\bm{y};\bm\theta\right)\\
  &-\mathbb{E}\left(\partial_{\bm\theta}\log p(\bm{y},\bm{\varphi};\bm\theta)|\bm{y};\bm\theta\right)\mathbb{E}\left(\partial_{\bm\theta}\log p(\bm{y},\bm{\varphi};\bm\theta)|\bm{y};\bm\theta\right)^{\top} 
\end{align*}
Given this, the second order derivative of the observed likelihood function with respect to parameter $\hat{\bm\theta}$, $\partial_{\bm\theta}^2 L(\hat{\bm\theta}; \bm y)$, can be approximated by the sequence $\{H_q\}_{q\in\mathbb{N}}$ which is calculated at iteration $q$ of the SAEM algorithm as:
\begin{eqnarray*}
    D_q&=&D_{q-1}+\gamma_q\left[ \partial_\theta \log p(\bm{y},\bm{\varphi}^{(q)};\bm{\theta}^{(q)} ) - D_{q-1} \right]\\
    G_q &=& G_{q-1}+\gamma_q\left[ \partial^2_\theta \log p(\bm{y},\bm{\varphi}^{(q)};\bm{\theta}^{(q)} ) \right.\\
    && \left. +\partial_\theta \log p(\bm{y},\bm{\varphi}^{(q)};\bm{\theta}^{(q)} )\partial_\theta \log p(\bm{y},\bm{\varphi}^{(q)};\bm{\theta}^{(q)} )^\prime- G_{q-1} \right]
\end{eqnarray*}
and $H_q = G_q-D_qD_q^\prime$. At convergence, $-H_q^{-1}$ can be used to approximate the covariance matrix of the parameter estimates \citep{biblio42,biblio43}, which are useful to derivate procedures for hypothesis testing for the different parameters of the model.

\section{Simulation studies}\label{sec3}

To evaluate the behavior of the proposed estimation method, and to compare it with existing alternatives, we conducted several simulation studies. It is worth noticing that the GHQ approach does not allow to deal with a different number of observations per individual, which is possible with our SAEM-based approach and the \texttt{gamlss} package. Therefore, we present simulations with balanced data first. Additional simulations on unbalanced data are also provided in Appendix A, in which the performance of SAEM on the unbalanced datasets is compared with the use of the GHQ algorithm on balanced datasets obtained from imputation, and with \texttt{gamlss} without imputation. Covariates significance analysis based on the LRT and the Wald test are also presented in Appendix B. To evaluate how the statistical power of the SAEM and GHQ methods varies with the proportion of zeros in the data, we performed a set of simulation experiments. The corresponding results are reported in Appendix C.

\subsection{Setup}\label{sec31}

We use two different settings for generating synthetic data under the ZIBR model (Equations \ref{zibr1} and \ref{zibr2}). The parameters for each configuration were chosen as follows:
\begin{itemize}
    \item \textit{Setting 1:} $a=b=-0.5$, $\alpha=\beta=0.5$, $\sigma_1=3.2$, $\sigma_2=2.6$, $\phi=6.4$.
    \item \textit{Setting 2:} $a=b=-0.5$, $\alpha=\beta=0.5$, $\sigma_1=0.7$, $\sigma_2=0.5$, $\phi=6.4$.
\end{itemize}
In the balanced scenario, for both Settings 1 and 2 the number of individuals $N=100$ will remain fixed, but the number of observations per individual $T_i$ will change, making $T_i=T$ with $T=3,5,10$. In addition, a variable $X$ is defined that mimics the concept of treatment and control groups, making $X=0$ for the first half of individuals and $X=1$ for the other half. Furthermore, we consider the same variable as covariate in both parts of the models, making $Z=X$.

For both Settings 1 and 2, $R=1000$ datasets were generated, and the SAEM estimation was implemented with $m=5$ chains and $(K_1,K_2)=(750,250)$, having therefore 1000 total iterations, with a starting point $\bm\theta_0=(\phi_0,a_0,b_0,\alpha_0,\beta_0,\sigma_{1,0},\sigma_{2,0})=(8,-0.3,-0.2,0.7,0.8,0.38,0.31)$. These quantities were determined using a graphical approach based on the trace plots of the parameter estimates across iterations. Convergence was assessed through visual inspection of the parameter trajectories. Initially, a large value of $K_1$ (with $\gamma_k$=1 for all iterations up to $K_1$) was used. From these plots, an appropriate value of $K_1$ was then selected once the iterates began to fluctuate randomly around a stable region.

\subsection{Results}\label{sec32}

\begin{figure*}[htbp]%
\centering
\includegraphics[width=\textwidth]{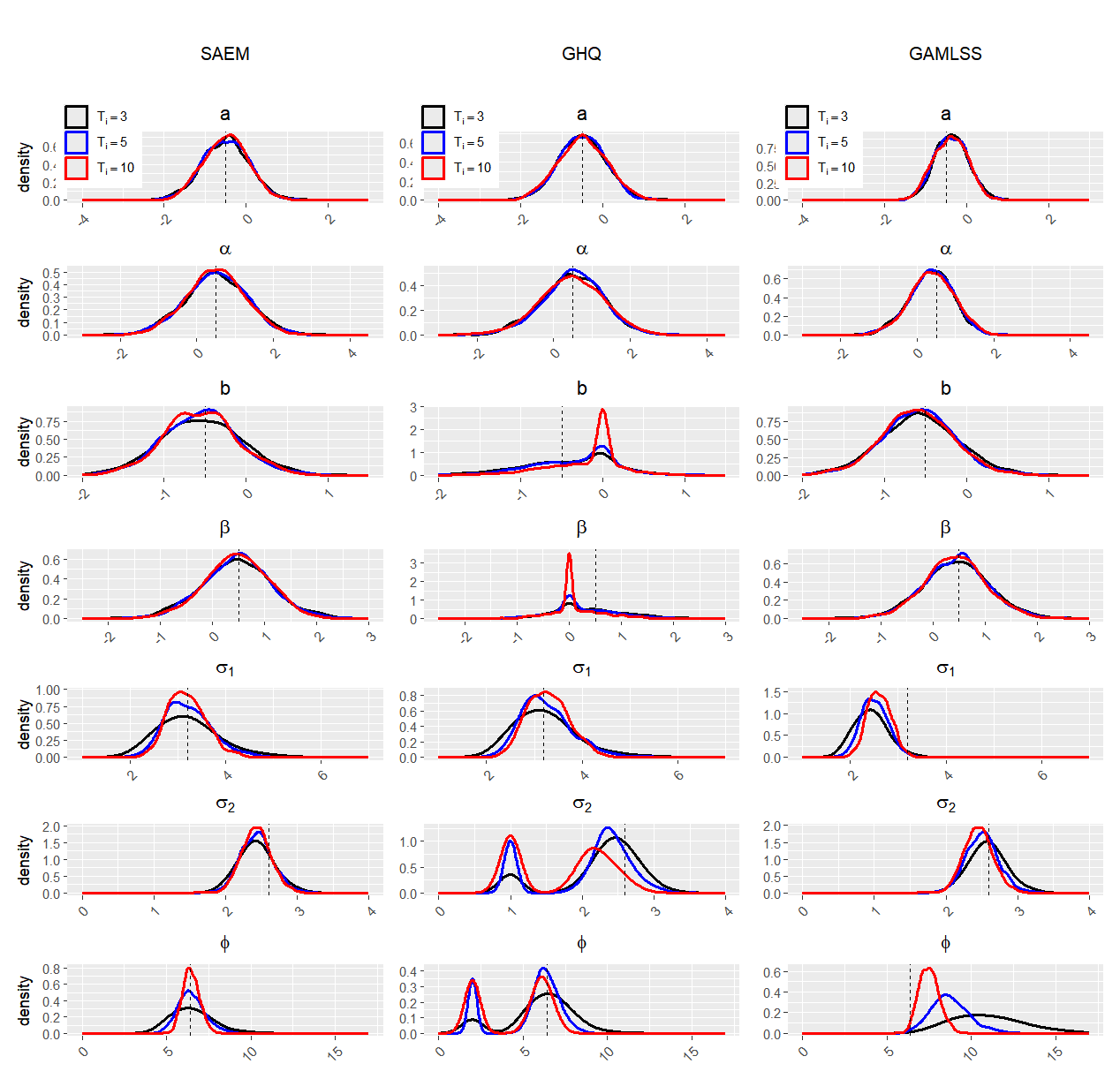}
\caption{Estimated density of the parameters obtained by the SAEM algorithm, the GHQ method and the GAMLSS procedure on artificial balanced datasets simulated under Setting 1. The dotted vertical line represents the true value of the parameter.}\label{fig1}
\end{figure*}

\begin{figure*}[htbp]%
\centering
\includegraphics[width=\textwidth]{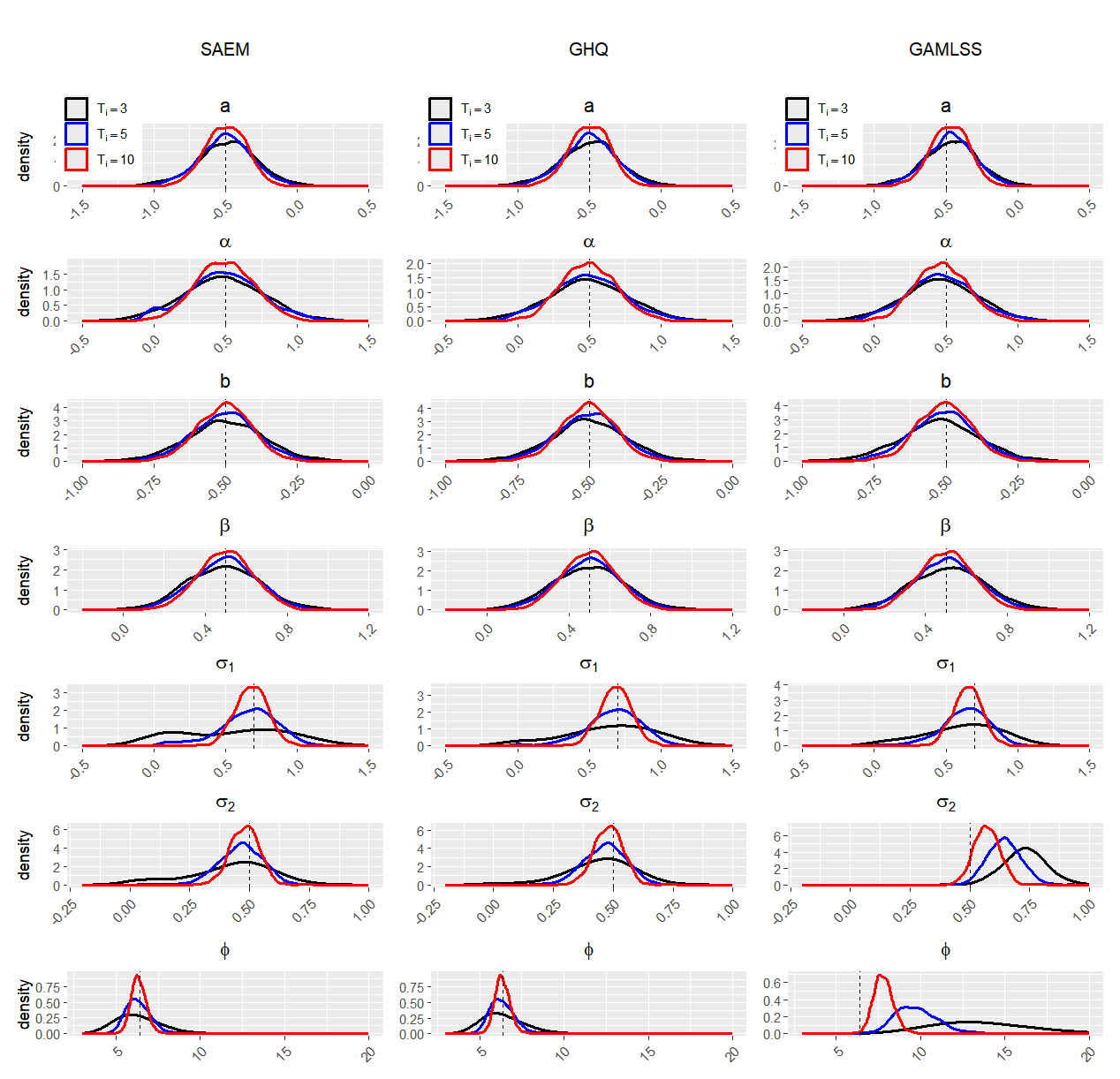}
\caption{Estimated density of the parameters obtained by the SAEM algorithm, the GHQ method and the GAMLSS  procedure on artificial balanced datasets simulated under Setting 2. The dotted vertical line represents the true value of the parameter.}\label{fig2}
\end{figure*}

Table \ref{tab1} shows the performance analysis of the two estimation methods for Settings 1 and 2 on balanced datasets, evaluated by bias $\left(\frac{1}{R}\sum_{r=1}^R \hat{\theta}^r - \theta\right)$, mean absolute error $\left(\mbox{MAE} = \frac{1}{R}\sum_{r=1}^R |\hat{\theta}^r - \theta|\right)$ and root mean square error $\left(\mbox{RMSE} = \sqrt{\frac{1}{R}\sum_{r=1}^R (\hat{\theta}^r - \theta)^2}\right)$. 

\begin{table}[htbp]
\hspace*{-2cm}
\caption{Summary statistics of the results obtained by SAEM algorithm, the GHQ procedure and the GAMLSS method on balanced data sets over 1000 simulation runs. For each parameter value and number of observations per individual, $T_i$, bold numbers indicate the lowest (absolute) value for each of bias, RMSE and MAE.}\label{tab1}
\hspace*{-5mm}
\begin{tabular}{lclccccccccc}
\toprule
\textbf{Parameter} & \textbf{Value} & & \textbf{Bias} & \textbf{RMSE} & \textbf{MAE} & \textbf{Bias} & \textbf{RMSE} & \textbf{MAE} & \textbf{Bias} & \textbf{RMSE} & \textbf{MAE}\\
\toprule

&&&\multicolumn{3}{c}{\textbf{SAEM}}&\multicolumn{3}{c}{\textbf{GHQ}}&\multicolumn{3}{c}{\textbf{GAMLSS}}\vspace*{-1mm}\\
\cmidrule{4-6}\cmidrule{7-9}\cmidrule{10-12}
$a$ & -0.5 & $T_i=3$ & 0.0064 & 0.5896 & 0.4629 & {\bf -0.0010} & 0.6509 & 0.4734 & 0.1460 & {\bf 0.4348} & {\bf 0.3449}\vspace*{-0.1mm}\\
 & & $T_i=5$ & 0.0081 & 0.5626 & 0.4526 & {\bf 0.0044} & 0.5634 & 0.4514 & 0.1302 & {\bf 0.4281} & {\bf 0.3436}\vspace*{-0.1mm}\\
 & & $T_i=10$ & {\bf 0.0154} & 0.5369 & 0.4306 & 0.0221 & 0.5947 & 0.4749 & 0.1139 & {\bf 0.4269} & {\bf 0.3469}\vspace*{-0.1mm}\\

$\alpha$ & 0.5 & $T_i=3$ & {\bf 0.0091} & 0.8277 & 0.6519 & 0.0215 & 0.9634 & 0.6587 & -0.1343 & {\bf 0.5846} & {\bf 0.4626}\vspace*{-0.1mm}\\
 & & $T_i=5$ & {\bf 0.0046} & 0.8001 & 0.6340 & 0.0138 & 0.7926 & 0.6186 & -0.1203 & {\bf 0.5822} & {\bf 0.4590}\vspace*{-0.1mm}\\
 & & $T_i=10$ & {\bf -0.0046} & 0.7533 & 0.6007 & -0.0253 & 0.8363 & 0.6625 & -0.1026 & {\bf 0.5827} & {\bf 0.4673}\vspace*{-0.1mm}\\

$b$ & -0.5 & $T_i=3$ & {\bf -0.0414} & 0.4936 & 0.3970 & 0.0970 & 0.5570 & 0.4607 & -0.0723 & {\bf 0.4744} & {\bf 0.3767}\vspace*{-0.1mm}\\
 & & $T_i=5$ & {\bf -0.0532} & 0.4511 & 0.3560 & 0.1752 & 0.5057 & 0.4317 & -0.0829 & {\bf 0.4416} & {\bf 0.3478}\vspace*{-0.1mm}\\
 & & $T_i=10$ & {\bf -0.0552} & 0.4392 & 0.3512 & 0.3197 & 0.5155 & 0.4582 & -0.0993 & {\bf 0.4375} & {\bf 0.3481}\vspace*{-0.1mm}\\

$\beta$ & 0.5 & $T_i=3$ & {\bf -0.0384} & 0.6827 & 0.5372 & -0.1442 & 0.7253 & 0.5788 & -0.0493 & {\bf 0.6529} & {\bf 0.5122} \vspace*{-0.1mm}\\
 & & $T_i=5$ & {\bf -0.0216} & 0.6530 & 0.5092 & -0.2532 & 0.6784 & 0.5544 & -0.0526 & {\bf 0.6160} & {\bf 0.4809}\vspace*{-0.1mm}\\
 & & $T_i=10$ & {\bf -0.0275} & 0.6083 & 0.4825 & -0.3828 & {\bf 0.5972} & 0.5202 & -0.0660 & 0.5980 & {\bf 0.4704}\vspace*{-0.1mm}\\

$\sigma_1$ & 3.2 & $T_i=3$ & {\bf 0.0284} & {\bf 0.6423} & {\bf 0.4961} & 0.0693 & 1.1306 & 0.5277 & -0.7702 & 0.8400 & 0.7754\vspace*{-0.1mm}\\
 & & $T_i=5$ & {\bf 0.0259} & {\bf 0.4920} & {\bf 0.3904} & 0.0631 & 0.5534 & 0.4238 & -0.6914 & 0.7461 & 0.6932\vspace*{-0.1mm}\\
 & & $T_i=10$ & {\bf -0.0012} & {\bf 0.3951} & {\bf 0.3170} & 0.0870 & 0.4433 & 0.3550 & -0.5939 & 0.6428 & 0.5956\vspace*{-0.1mm}\\

$\sigma_2$ & 2.6 & $T_i=3$ & -0.1720 & 0.2979 & 0.2429 & -0.3206 & 0.6612 & 0.4347 & {\bf -0.0142} & {\bf 0.2478} & {\bf 0.1957}\vspace*{-0.1mm}\\
 & & $T_i=5$ & -0.1639 & 0.2779 & 0.2270 & -0.4970 & 0.8065 & 0.5605 & {\bf -0.0899} & {\bf 0.2425} & {\bf 0.1932}\vspace*{-0.1mm}\\
 & & $T_i=10$ & -0.1699 & 0.2635 & 0.2162 & -0.8565 & 1.0651 & 0.8674 & {\bf -0.1457} & {\bf 0.2480} & {\bf 0.2018}\vspace*{-0.1mm}\\

$\phi$ & 6.4 & $T_i=3$ & {\bf 0.1301} & {\bf 1.2251} & {\bf 0.9465} & -0.3696 & 2.0104 & 1.4482 & 4.6834 & 5.1779 & 4.6834\vspace*{-0.1mm}\\
 & & $T_i=5$ & {\bf 0.0895} & {\bf 0.8059} & {\bf 0.6283} & -0.9741 & 2.1842 & 1.4390 & 2.3698 & 2.6248 & 2.3713\vspace*{-0.1mm}\\
 & & $T_i=10$ & {\bf 0.0645} & {\bf 0.4940} & {\bf 0.3931} & -1.9106 & 2.8099 & 2.0202 & 1.1085 & 1.2564 & 1.1139\vspace*{0.5mm}\\

&&&\multicolumn{3}{c}{\textbf{SAEM}}&\multicolumn{3}{c}{\textbf{GHQ}}&\multicolumn{3}{c}{\textbf{GAMLSS}}\vspace*{-1mm}\\
\cmidrule{4-6}\cmidrule{7-9}\cmidrule{10-12}

$a$ & -0.5 & $T_i=3$ & 0.0106 & 0.2101 & 0.1659 & {\bf 0.0026} & 0.2060 & 0.1623 & 0.0376 & {\bf 0.1935} & {\bf 0.1540}\vspace*{-0.1mm}\\
 & & $T_i=5$ & 0.0017 & 0.1813 & 0.1432 & {\bf -0.0003} & 0.1780 & 0.1404 & 0.0317 & {\bf 0.1688} & {\bf 0.1340}\vspace*{-0.1mm}\\
 & & $T_i=10$ & 0.0033 & 0.1414 & 0.1144 & {\bf 0.0023} & 0.1380 & 0.1114 & 0.0276 & {\bf 0.1337} & {\bf 0.1080}\vspace*{-0.1mm}\\

$\alpha$ & 0.5 & $T_i=3$ & -0.0120 & 0.2912 & 0.2283 & {\bf -0.0040} & 0.2859 & 0.2243 & -0.0385 & {\bf 0.2663} & {\bf 0.2102}\vspace*{-0.1mm}\\
 & & $T_i=5$ & -0.0060 & 0.2551 & 0.2025 & {\bf -0.0032} & 0.2482 & 0.1973 & -0.0349 & {\bf 0.2346} & {\bf 0.1870}\vspace*{-0.1mm}\\
 & & $T_i=10$ & -0.0042 & 0.1997 & 0.1607 & {\bf -0.0029} & 0.1910 & 0.1528 & -0.0282 & {\bf 0.1833} & {\bf 0.1468}\vspace*{-0.1mm}\\

$b$ & -0.5 & $T_i=3$ & {\bf 0.0006} & 0.1348 & 0.1063 & -0.0025 & {\bf 0.1331} & {\bf 0.1043} & -0.0212 & 0.1409 & 0.1102\vspace*{-0.1mm}\\
 & & $T_i=5$ & -0.0031 & 0.1099 & {\bf 0.0873} & {\bf -0.0028} & {\bf 0.1096} & 0.0874 & -0.0094 & 0.1116 & 0.0885\vspace*{-0.1mm}\\
 & &  $T_i=10$ & -0.0032 & {\bf 0.0922} & 0.0732 & {\bf -0.0031} & {\bf 0.0922} & {\bf 0.0729} & -0.0040 & 0.0928 & 0.0737\vspace*{-0.1mm}\\

$\beta$ & 0.5 & $T_i=3$ & -0.0089 & 0.1792 & 0.1430 & {\bf -0.0049} & {\bf 0.1753} & {\bf 0.1403} & 0.0122 & 0.1829 & 0.1463\vspace*{-0.1mm}\\
 & & $T_i=5$ & {\bf -0.0006} & 0.1536 & 0.1223 & -0.0011 & {\bf 0.1516} & {\bf 0.1206} & 0.0050 & 0.1542 & 0.1226\vspace*{-0.1mm}\\
 & & $T_i=10$ & 0.0015 & 0.1326 & 0.1060 & {\bf 0.0012} & {\bf 0.1308} & {\bf 0.1044} & {\bf 0.0012} & 0.1314 & 0.1052\vspace*{-0.1mm}\\

$\sigma_1$ & 0.7 & $T_i=3$ & -0.1448 & 0.3944 & 0.3239 & {\bf -0.0535} & 0.3176 & 0.2501 & -0.0845 & {\bf 0.2768} & {\bf 0.2164}\vspace*{-0.1mm}\\
 & & $T_i=5$ & -0.0394 & 0.2107 & 0.1605 & {\bf -0.0275} & 0.1951 & 0.1499 & -0.0643 & {\bf 0.1808} & {\bf 0.1373}\vspace*{-0.1mm}\\
 & & $T_i=10$ & -0.0192 & 0.1137 & 0.0910 & {\bf -0.0169} & 0.1108 & 0.0883 & -0.0489 & {\bf 0.1106} & {\bf 0.0877}\vspace*{-0.1mm}\\

$\sigma_2$ & 0.5 & $T_i=3$ & -0.0812 & 0.1918 & 0.1407 & {\bf -0.0542} & {\bf 0.1529} & {\bf 0.1139} & 0.2324 & 0.2481 & 0.2328\vspace*{-0.1mm}\\
 & & $T_i=5$ & -0.0359 & 0.0991 & 0.0771 & {\bf -0.0337} & {\bf 0.0963} & {\bf 0.0749} & 0.1453 & 0.1610 & 0.1460\vspace*{-0.1mm}\\
 & & $T_i=10$ & -0.0193 & 0.0629 & 0.0499 & {\bf -0.0190} & {\bf 0.0624} & {\bf 0.0496} & 0.0715 & 0.0891 & 0.0754\vspace*{-0.1mm}\\

$\phi$ & 6.4 & $T_i=3$ & -0.0920 & 1.2570 & 1.0054 & {\bf 0.0031} & {\bf 1.1890} & {\bf 0.9402} & 7.4381 & 8.0589 & 7.4381\vspace*{-0.1mm}\\
 & & $T_i=5$ & -0.0700 & 0.7283 & 0.5820 & {\bf -0.0620} & {\bf 0.7214} & {\bf 0.5767} & 3.3718 & 3.6059 & 3.3718\vspace*{-0.1mm}\\
 & & $T_i=10$ & {\bf -0.0463} & 0.4423 & 0.3518 & -0.0472 & {\bf 0.4409} & {\bf 0.3510} & 1.3984 & 1.5107 & 1.4002\vspace*{0.5mm}\\

\bottomrule
\end{tabular}
\end{table}

Analyzing these results globally, and the global estimates distributions in Figures \ref{fig1} and \ref{fig2}, we can see how SAEM estimates are always centered (except for a small bias for $\sigma_2$ in Setting 1), whereas GHQ and GAMLSS methods present strong biases or bimodality for different parameters: $b$, $\beta$, $\sigma_2$ and $\phi$ for GHQ in Setting 1 and  $\sigma_1$ (only Setting 1), $\sigma_2$ and $\phi$ (both settings) for GAMLSS. 
A thorough examination of these results reveals that the SAEM estimation achieves optimal bias performance in Setting 1. In Setting 2, the GHQ method exhibits slightly lower bias values. It merits attention, however, that GAMLSS attains the lowest root mean square error (RMSE) and mean absolute error (MAE) values in many parameters across both settings. It is worth noticing, nevertheless, that the GAMLSS estimates for $\phi$ are extremely biased in both settings. These two situations are common in estimators obtained by quasi-likelihood methods, as indicated by \cite{biblio57}. 

A remarkable property of SAEM is its tendency to exhibit a consistent decrease in the error measures as the number of observations per individual increases, a phenomenon not consistently observed in the other analyzed alternatives. In Setting 1, where the values of the variance parameters are higher, it is evident that the worst GHQ results are obtained in the estimation of the parameters associated with the Beta part of the model ($b, \beta, \sigma_2$ and $\phi$). This component has a complex functional form that could be incorrectly approximated when integrating by numerical methods. The SAEM approach is advantageous in this regard. In the context of GAMLSS, the estimation of the parameter controlling the overdispersion of the data, i.e., $\phi$, is particularly challenging (as well as the estimation of $\sigma_2$ in Setting 2). According to \cite{biblio56}, by defining ZIBR as a model with several random effects, GAMLSS estimates the variance components with a method prone to generating biases. Given that overdispersion is a common feature in real microbiota data, the GAMLSS procedure may not be suitable for modeling this phenomenon. 

A more detailed analysis of the Figures shows that for Setting 1 the estimated densities of $a$ and $\alpha$ are very similar in all methods, while the other parameters show marked differences. In the case of GHQ, the results for $b$ and $\beta$ are very skewed, while the distribution of $\sigma_2$ and $\phi$ shows bimodal behavior. This is due to the poor approximation of the log-likelihood, which leads to an erroneous estimation of these parameters. A review of the GAMLSS results reveals that the distribution of $\sigma_1$ is significantly deviated from the true parameter, while the distribution of $\phi$ exhibits considerable variability when observations per individual are limited. Figure \ref{fig2}, on the other hand, shows that for Setting 2 SAEM and GHQ are practically equivalent in the estimates densities, while GAMLSS differs only in that it performs poorly for $\sigma_2$ and $\phi$. This indicates that the minor errors differences between the methods in favor of GHQ for this simulation scenario (Table \ref{tab1}) are not significant. 

Regarding the execution times of the routines, GHQ and GAMLSS are faster than SAEM in the 3 cases considered in both Settings. On a computer with Intel Core i7-13700HX processor at 2.10 GHz, for $T_i=3\ (5, 10)$, on average, GAMLSS takes $0.94\  (1.03, 1.20)$ seconds, GHQ takes $1.77\  (3.39, 8.54)$ seconds,  while SAEM takes $9.86\ (14.14, 25.30)$ seconds.

\section{Case studies}\label{sec4}

In this section we demonstrate the use of the proposed inference framework on a publicly available microbiome study and we will focus on the capabilities of the SAEM algorithm to detect changes in the presence of bacterial taxa in response to treatments. The study of another set of microbiome data can be found in Appendix D of the supplementary materials. As we mentioned in the previous section, given that the ZIBR model has more than one random effect, GAMLSS uses an estimation method that can cause a bias in the estimation of the variance parameters, and also does not allow the evaluation of the tests based on the calculation of the log-likelihood that we will use in the following section. For this reason, the GAMLSS approach will not be applied to the real data considered below.

\subsection{Inflammatory bowel disorder pediatric study}\label{sec41}

\begin{figure}[htb]%
\centering
\includegraphics[width=\textwidth]{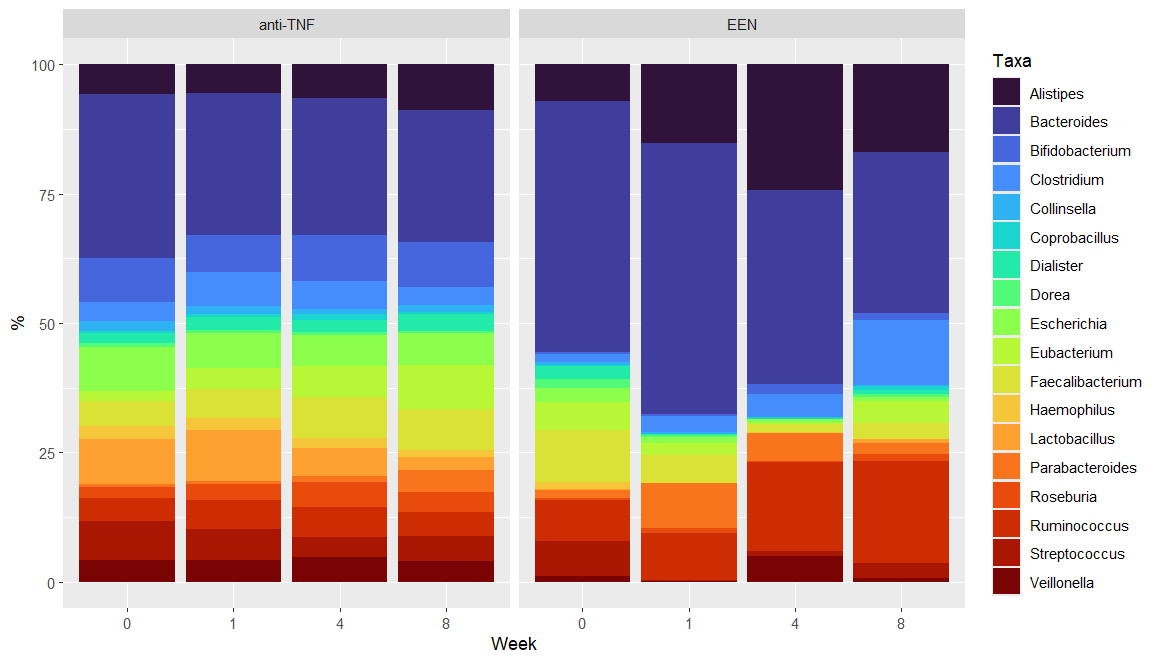}
\caption{Average gut microbiome composition of the treatment groups (anti-TNF and EEN) over observation week}\label{fig3}
\end{figure}

The data used in this section come from a study to verify the effectiveness of treatments in pediatric inflammatory bowel disorder (IBD) patients \citep{biblio24}. This study includes information from 90 children subjected to three different types of therapy: anti-TNF treatment (TNF: tumor necrosis factor), exclusive enteral nutrition (EEN) and partial enteral nutrition with an ad lib diet (PEN). After filtering the data to discard low sequencing depth samples, low abundant genus and taxa with a proportion of zeros higher than 0.9 or lower than 0.1, the information from 18 bacterial genera measured at 4 different time points for each one of the 59 individuals (47 anti-TNF and 12 EEN) remained for the analysis. Figure \ref{fig3} shows the average composition of the intestinal microbiome of the subjects in both groups and its evolution over time.

The objective of the study is to verify if the different treatments influence the presence of the different bacterial taxa in the samples, controlling for time and initial abundance. In addition, we want to compare if the results obtained by the SAEM algorithm differ from those obtained through the GHQ procedure implemented in the \texttt{ZIBR} package. The initial values for SAEM were the estimates found by the GHQ method, for each model corresponding to each bacterial taxon. Given this choice of initial values, we used $m=5$ Markov chains and $(K_1,K_2)=(375,125)$ iterations, as well as 500 simulated values for the log-likelihood calculation by Importance Sampling. The p-values obtained through the LRT were adjusted using the Benjamini-Hochberg process \citep{biblio30} to decrease the false discovery rate (FDR) and the full values are presented in Table 8, Appendix D of Supplementary Materials.

\begin{figure}[htb]%
\centering
\includegraphics[width=0.7\textwidth]{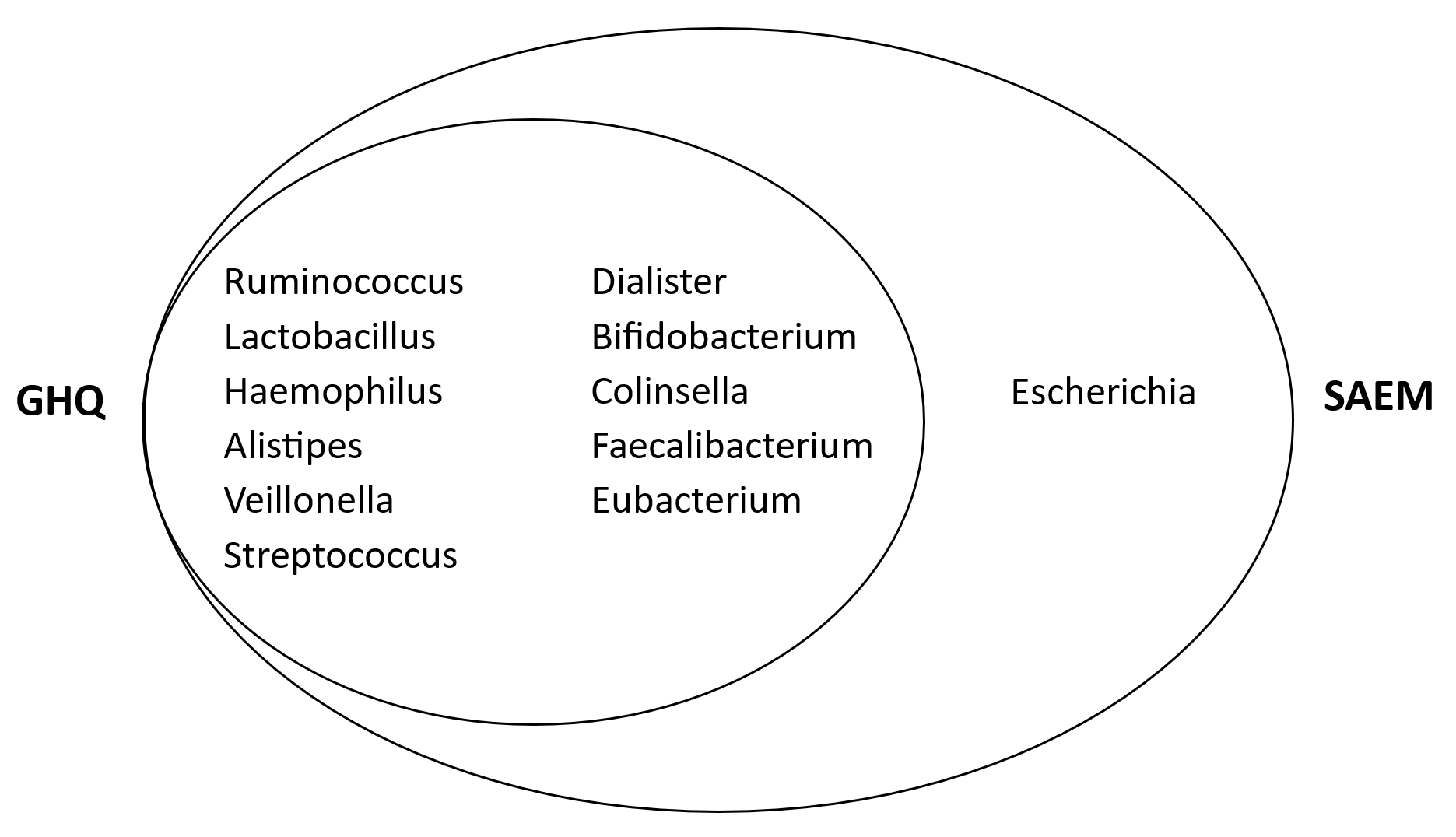}
\caption{Bacterial taxa in which the treatment (anti-TNF vs. EEN) have a statistical effect in abundance identified by SAEM and GHQ}\label{fig4}
\end{figure}

After model fitting, the GHQ method detected 11 bacterial taxa in which the treatment influenced the abundances, while SAEM managed to identify 12 taxa, all those identified by GHQ in addition to \textit{Escherichia} with FDR=5\%, as shown in Figure \ref{fig4}.

A more detailed analysis of the \textit{Escherichia} data shows that the influence of treatment is greater on the frequency of presence of this bacterium in individuals than on the level of abundance once its presence is confirmed (Figure 5 in Appendix D). This is confirmed by the LRT results for the significance of the treatment in the calculation of the probability $p_{it}$ (FDR p-value 0.03) compared to those of the Beta component of the abundance $u_{it}$ (FDR p-value 0.80), and by the Wald test (Table 9, Appendix D). These results show that at the 5\% significance level the treatment is significant in the logistic part but not in the Beta part, proving that the definition of the ZIBR model and the combination with the SAEM estimation allows the increase of the ability to detect the influence of a given treatment defined. A detailed figure with the convergence behavior of the estimators across iterations is shown in Figure 6, Appendix D of the supplementary materials.

The role of \textit{Escherichia} in IBD is well documented. There is evidence \citep{biblio40} that the accumulation of \textit{Escherichia coli} and other strains of \textit{Escherichia} in the intestine is related to inflammatory processes, and other works \citep{biblio39} suggest that a combination of antibiotic and dietary treatments is capable of controlling overproliferation of \textit{E. coli} in the digestive system and also reducing the symptoms of IBD, allowing to infer a correlation between these two events.

\section{Conclusions and discussion}\label{sec5}

In this article we have developed an exact maximum likelihood estimation strategy for the ZIBR model using the SAEM algorithm. We have also proposed a method for calculating the log-likelihood of the model which allows to obtain information criteria for the model, and approximations of the estimators standard errors, which is not possible under the alternative estimation method based on Gauss-Hermite quadrature likelihood approximation. Throughout the article, the analysis of longitudinal compositional microbiome data motivates the specific characteristics of this model and the need for more efficient estimation methods. Nonetheless, the ZIBR model and the methodology developed here can be used in a wide range of applications in which longitudinal proportion data exhibit a large number of zeros.

Compared with the GAMLSS alternative, which can also estimate ZIBR parameters and standard errors, SAEM shows superior control of type I error in Wald tests. Although the results obtained by SAEM conform to the expected theoretical properties of standard errors, it must be noted that the method can still be improved, since, as it depends on a stochastic approximation, convergence towards coherent values for the Fisher information matrix is not guaranteed, which could introduce bias in the estimation of the standard errors. We are confident that these details can be enhanced in further developments.

A key advantage of the SAEM-based method is its ability to handle unbalanced data. This is essential in clinical studies, where dropout or irregular follow-up is frequent; yet the original ZIBR estimation method cannot accommodate such designs without prior interpolation. The GAMLSS method, another option that allows for handling unbalanced data, has been found to show errors in the estimation of certain important parameters of the model, even when its performance in the rest is adequate. It should be emphasized that unbalanced data is a fairly common situation in medical experiments, in which multiple factors influence patients abandoning the follow-up. This could be one of the reasons contributing to the high non-publication rate in many medical studies, which according to certain sources could be close to 50\% \citep{biblio35}. Therefore, developing analysis methods that can deal accurately with unbalanced data is of great interest.

The definition of the ZIBR model used throughout this work corresponds to the one originally proposed by \cite{biblio9} and implemented in the \texttt{ZIBR} package for R statistical software. However, there are possibilities for modification of this definition that have been discussed. One of them is the use of random effects for more than one covariate, an aspect that has been already incorporated in the implementation used in this article. Another possibility is the inclusion of cross correlations in the random effects, proposing a different structure in the variance of these effects. \cite{biblio36} mention that this inclusion could alter the results for tests on covariates, detecting significance where a simpler structure would not detect it. Although this approach has not been implemented here, the SAEM algorithm could be easily modified to serve this purpose.

It is important to notice that our approach focuses on marginal taxon-wise modeling. When the goal is to study the joint evolution of taxa distributions, multivariate models should be considered. Several alternatives to ZIBR have been proposed in microbiome research. Among the most important are ZIBR-SRE \citep{biblio37}, an extension of ZIBR which considers the compositional nature of microbiota data; zero-inflated Gaussian mixed models (ZIGMM) \citep{biblio16}, which in addition to managing the overabundance of zeros can work with both proportion data and counts; and the negative binomial mixed model (NBMM) \citep{biblio38}, which allows the specification of more general variance structures and also be modified to deal with zero inflation. It seems interesting to implement the SAEM algorithm to these models and study its potential benefits in estimation.

Finally, an interesting extension of this work would be to obtain the Restricted Maximum Likelihood (REML) estimates, a known method for reducing bias in the estimation of variance components in mixed effects models \citep{biblio25}, using the Harville's approach, i.e., integrating out the fixed effects, via the SAEM algorithm. We expect that, in the context of longitudinal models on microbiome data, this could improve the results obtained through ML estimation.

\section*{Supplementary materials}

The supplementary materials include extended simulation studies, tables and figures referenced in Sections \ref{sec3} and \ref{sec4} and an additional case study. R codes to implement the routines of the SAEM estimation and to reproduce the statistical analysis of Section \ref{sec4} are available at \url{https://github.com/jbarrera232/saem-zibr}.

\section*{Declaration of conflicting interests}
The authors declared no potential conflicts of interest with respect to the research, authorship and/or publication of this article.

\section*{Funding}
The work of the first and second author was supported by ANID MATH-AmSud Project AMSUD 230032-SMILE. The work of the first author was also supported by ANID Becas/Doctorado Nacional 21231659. The third author gratefully acknowledges the support of grants PID2021-123592OB-I00 funded by MCIN/AEI/10.13039 /501100011033 and by ERDF - A way of making Europe, and TED2021-129316B-I00 funded by MCIN/AEI/10.13039 /501100011033 and by the European Union NextGenerationEU/PRTR.


\begin{thebibliography}{}

\bibitem[Arribas-Gil et~al., 2014]{ABMR:2014}
Arribas-Gil, A., Bertin, K., Meza, C., and Rivoirard, V. (2014).
\newblock {LASSO}-type estimators for semiparametric nonlinear mixed-effects models estimation.
\newblock {\em Statistics and Computing}, 24(3):443--460.

\bibitem[Baldelli et~al., 2021]{biblio40}
Baldelli, V., Scaldaferri, F., Putignani, L., and Del~Chierico, F. (2021).
\newblock The role of {Enterobacteriaceae} in gut microbiota dysbiosis in inflammatory bowel diseases.
\newblock {\em Microorganisms}, 9(4):697.

\bibitem[Benjamini and Hochberg, 1995]{biblio30}
Benjamini, Y. and Hochberg, Y. (1995).
\newblock Controlling the false discovery rate: a practical and powerful approach to multiple testing.
\newblock {\em Journal of the Royal Statistical Society: series B (Methodological)}, 57(1):289--300.

\bibitem[Cai, 2010]{biblio43}
Cai, L. (2010).
\newblock High-dimensional exploratory item factor analysis by a {Metropolis-Hastings} {Robbins-Monro} algorithm.
\newblock {\em Psychometrika}, 75(1):33--57.

\bibitem[Celeux et~al., 1995]{biblio50}
Celeux, G., Chauveau, D., and Diebolt, J. (1995).
\newblock {On Stochastic Versions of the EM Algorithm}.
\newblock Research Report RR-2514, {INRIA}.

\bibitem[Chan et~al., 2014]{biblio35}
Chan, A.~W., Song, F., Vickers, A., Jefferson, T., Dickersin, K., G{\o}tzsche, P.~C., Krumholz, H.~M., Ghersi, D., and Van Der~Worp, H.~B. (2014).
\newblock Increasing value and reducing waste: addressing inaccessible research.
\newblock {\em The Lancet}, 383(9913):257--266.

\bibitem[Chen and Li, 2016]{biblio9}
Chen, E.~Z. and Li, H. (2016).
\newblock A two-part mixed-effects model for analyzing longitudinal microbiome compositional data.
\newblock {\em Bioinformatics}, 32(17):2611--2617.

\bibitem[Clemente et~al., 2012]{biblio4}
Clemente, J.~C., Ursell, L.~K., Parfrey, L.~W., and Knight, R. (2012).
\newblock The impact of the gut microbiota on human health: an integrative view.
\newblock {\em Cell}, 148(6):1258--1270.

\bibitem[Comets et~al., 2021]{saemix_userguide}
Comets, E., Karimi, B., Delattre, M., Ranke, J., Lavenu, A., Lavielle, M., Chanel, M., Guhl, M., Fayette, L., and Kaisaridi, S. (2021).
\newblock Saemix user's guide, version 3.0.
\newblock \url{https://github.com/iame-researchCenter/saemix/blob/7638e1b09ccb01cdff173068e01c266e906f76eb/docsaem.pdf}.

\bibitem[Comets et~al., 2017]{saemix}
Comets, E., Lavenu, A., and Lavielle, M. (2017).
\newblock Parameter estimation in nonlinear mixed effect models using saemix, an {R} implementation of the {SAEM} algorithm.
\newblock {\em Journal of Statistical Software}, 80(3):1--41.

\bibitem[{de la Cruz} et~al., 2024]{DLMN:2024}
{de la Cruz}, R., Lavielle, M., Meza, C., and Núñez-Antón, V. (2024).
\newblock A joint analysis proposal of nonlinear longitudinal and time-to-event right-, interval-censored data for modeling pregnancy miscarriage.
\newblock {\em Computers in Biology and Medicine}, 182:109186.

\bibitem[Dekaboruah et~al., 2020]{biblio1}
Dekaboruah, E., Suryavanshi, M.~V., Chettri, D., and Verma, A.~K. (2020).
\newblock Human microbiome: an academic update on human body site specific surveillance and its possible role.
\newblock {\em Archives of {Microbiology}}, 202:2147--2167.

\bibitem[Delyon et~al., 1999]{biblio20}
Delyon, B., Lavielle, M., and Moulines, E. (1999).
\newblock Convergence of a stochastic approximation version of the {EM} algorithm.
\newblock {\em Annals of Statistics}, pages 94--128.

\bibitem[Dempster et~al., 1977]{biblio18}
Dempster, A.~P., Laird, N.~M., and Rubin, D.~B. (1977).
\newblock Maximum likelihood from incomplete data via the {EM} algorithm.
\newblock {\em Journal of the Royal Statistical Society: Series B (Methodological)}, 39(1):1--22.

\bibitem[D’Agata et~al., 2019]{biblio11}
D’Agata, A.~L., Wu, J., Welandawe, M. K.~V., Dutra, S. V.~O., Kane, B., and Groer, M.~W. (2019).
\newblock Effects of early life {NICU} stress on the developing gut microbiome.
\newblock {\em Developmental Psychobiology}, 61(5):650--660.

\bibitem[Eggers, 2015]{biblio48}
Eggers, J. (2015).
\newblock {\em On Statistical Methods for Zero-Inflated Models}.
\newblock Thesis, Uppsala Universitet.

\bibitem[Han et~al., 2021]{biblio37}
Han, Y., Baker, C., Vogtmann, E., Hua, X., Shi, J., and Liu, D. (2021).
\newblock Modeling longitudinal microbiome compositional data: a two-part linear mixed model with shared random effects.
\newblock {\em Statistics in Biosciences}, 13:243--266.

\bibitem[Handayani et~al., 2017]{biblio13}
Handayani, D., Notodiputro, K.~A., Sadik, K., and Kurnia, A. (2017).
\newblock A comparative study of approximation methods for maximum likelihood estimation in generalized linear mixed models {(GLMM)}.
\newblock {\em AIP Conference Proceedings}, 1827(1):020033.

\bibitem[Hu et~al., 2022]{biblio10}
Hu, J., Wang, C., Blaser, M.~J., and Li, H. (2022).
\newblock Joint modeling of zero-inflated longitudinal proportions and time-to-event data with application to a gut microbiome study.
\newblock {\em Biometrics}, 78(4):1686--1698.

\bibitem[Jeyakumar et~al., 2019]{biblio5}
Jeyakumar, T., Beauchemin, N., and Gros, P. (2019).
\newblock Impact of the microbiome on the human genome.
\newblock {\em Trends in Parasitology}, 35(10):809--821.

\bibitem[Kloek and van Dijk, 1978]{Kloek:1978}
Kloek, T. and van Dijk, H.~K. (1978).
\newblock Bayesian estimates of equation system parameters: An application of integration by {Monte Carlo}.
\newblock {\em Econometrica}, 46(1):1--19.

\bibitem[Kodikara et~al., 2022]{biblio8}
Kodikara, S., Ellul, S., and L{\^e}~Cao, K.-A. (2022).
\newblock Statistical challenges in longitudinal microbiome data analysis.
\newblock {\em Briefings in Bioinformatics}, 23(4):1--18.

\bibitem[Kuhn and Lavielle, 2005]{biblio23}
Kuhn, E. and Lavielle, M. (2005).
\newblock Maximum likelihood estimation in nonlinear mixed effects models.
\newblock {\em Computational Statistics \& Data Analysis}, 49(4):1020--1038.

\bibitem[Lewis et~al., 2015]{biblio24}
Lewis, J.~D., Chen, E.~Z., Baldassano, R.~N., Otley, A.~R., Griffiths, A.~M., Lee, D., Bittinger, K., Bailey, A., Friedman, E.~S., Hoffmann, C., et~al. (2015).
\newblock Inflammation, antibiotics, and diet as environmental stressors of the gut microbiome in pediatric {Crohn’s} disease.
\newblock {\em Cell Host \& Microbe}, 18(4):489--500.

\bibitem[Liu et~al., 2019]{biblio36}
Liu, L., Shih, Y.-C.~T., Strawderman, R.~L., Zhang, D., Johnson, B.~A., and Chai, H. (2019).
\newblock {Statistical Analysis of Zero-Inflated Nonnegative Continuous Data: A Review}.
\newblock {\em Statistical Science}, 34(2):253 -- 279.

\bibitem[Louis, 1982]{biblio41}
Louis, T. (1982).
\newblock Finding the observed information matrix when using the {EM} algorithm.
\newblock {\em Journal of the Royal Statistical Society. Series B (Methodological)}, 44(2):226--233.

\bibitem[McLachlan and Krishnan, 2008]{McLachlanKrishnan2008}
McLachlan, G.~J. and Krishnan, T. (2008).
\newblock {\em The EM Algorithm and Extensions}.
\newblock Wiley Series in Probability and Statistics. John Wiley \& Sons, Hoboken, NJ, 2nd edition.

\bibitem[Metropolis et~al., 1953]{biblio31}
Metropolis, N., Rosenbluth, A.~W., Rosenbluth, M.~N., Teller, A.~H., and Teller, E. (1953).
\newblock Equation of state calculations by fast computing machines.
\newblock {\em The Journal of Chemical Physics}, 21(6):1087--1092.

\bibitem[Meza et~al., 2007]{biblio25}
Meza, C., Jaffr{\'e}zic, F., and Foulley, J.-L. (2007).
\newblock {REML} estimation of variance parameters in nonlinear mixed effects models using the {SAEM} algorithm.
\newblock {\em Biometrical Journal: Journal of Mathematical Methods in Biosciences}, 49(6):876--888.

\bibitem[Meza et~al., 2012]{MOD:2012}
Meza, C., Osorio, F., and De~la Cruz, R. (2012).
\newblock Estimation in nonlinear mixed-effects models using heavy-tailed distributions.
\newblock {\em Statistics and Computing}, 22(1):121--139.

\bibitem[Min and Agresti, 2005]{biblio26}
Min, Y. and Agresti, A. (2005).
\newblock Random effect models for repeated measures of zero-inflated count data.
\newblock {\em Statistical Modelling}, 5(1):1--19.

\bibitem[Mirsepasi-Lauridsen et~al., 2019]{biblio39}
Mirsepasi-Lauridsen, H.~C., Vallance, B.~A., Krogfelt, K.~A., and Petersen, A.~M. (2019).
\newblock Escherichia coli pathobionts associated with inflammatory bowel disease.
\newblock {\em Clinical Microbiology Reviews}, 32(2):e00060--18.

\bibitem[Myers, 2000]{biblio15}
Myers, W.~R. (2000).
\newblock Handling missing data in clinical trials: an overview.
\newblock {\em Drug information journal: DIJ/Drug Information Association}, 34:525--533.

\bibitem[Márquez et~al., 2023]{biblio21}
Márquez, M., Meza, C., Lee, D.-J., and De~la Cruz, R. (2023).
\newblock Classification of longitudinal profiles using semi-parametric nonlinear mixed models with {P-splines} and the {SAEM} algorithm.
\newblock {\em Statistics in Medicine}, 42(27):4952--4971.

\bibitem[Nelder and Lee, 1992]{biblio57}
Nelder, J. and Lee, Y. (1992).
\newblock Likelihood, quasi-likelihood and pseudolikelihood: some comparisons.
\newblock {\em Journal of the Royal Statistical Society Series B: Statistical Methodology}, 54(1):273--284.

\bibitem[Ocaña-Riola et~al., 2021]{biblio62}
Ocaña-Riola, R., Pérez-Romero, C., Ortega-Díaz, M.~I., and Martín-Martín, J.~J. (2021).
\newblock Multilevel zero-one inflated beta regression model for the analysis of the relationship between exogenous health variables and technical efficiency in the {Spanish National Health System Hospitals}.
\newblock {\em International Journal of Environmental Research and Public Health}, 18(19).

\bibitem[Ospina and Ferrari, 2012]{biblio55}
Ospina, R. and Ferrari, S.~L. (2012).
\newblock A general class of zero-or-one inflated beta regression models.
\newblock {\em Computational Statistics \& Data Analysis}, 56(6):1609--1623.

\bibitem[Powney et~al., 2014]{biblio14}
Powney, M., Williamson, P., Kirkham, J., and Kolamunnage-Dona, R. (2014).
\newblock A review of the handling of missing longitudinal outcome data in clinical trials.
\newblock {\em Trials}, 15(1):1--11.

\bibitem[Rigby and Stasinopoulos, 1996]{biblio53}
Rigby, R.~A. and Stasinopoulos, D. (1996).
\newblock A semi-parametric additive model for variance heterogeneity.
\newblock {\em Statistics and Computing}, 6:57--65.

\bibitem[Rigby and Stasinopoulos, 2005]{biblio52}
Rigby, R.~A. and Stasinopoulos, D. (2005).
\newblock Generalized additive models for location, scale and shape, (with discussion).
\newblock {\em Applied Statistics}, 54:507--554.

\bibitem[Samson et~al., 2007]{biblio22}
Samson, A., Lavielle, M., and Mentr{\'e}, F. (2007).
\newblock The {SAEM} algorithm for group comparison tests in longitudinal data analysis based on non-linear mixed effects model.
\newblock {\em Statistics in Medicine}, 26(27):4860--4875.

\bibitem[Stasinopoulos et~al., 2024]{GAMLSS}
Stasinopoulos, M.~D., Kneib, T., Klein, N., Mayr, A., and Helle, G.~Z. (2024).
\newblock {\em Generalized Additive Models for Location, Scale and Shape: A Distributional Regression Approach, with Applications}.
\newblock Cambridge University Press.

\bibitem[Stasinopoulos et~al., 2017]{biblio56}
Stasinopoulos, M.~D., Rigby, R.~A., Heller, G.~Z., Voudouris, V., and De~Bastiani, F. (2017).
\newblock {\em Flexible regression and smoothing: using GAMLSS in R}.
\newblock CRC Press, Taylor \& Francis Group.

\bibitem[Tang et~al., 2023]{biblio63}
Tang, B., Frye, H.~A., Gelfand, A.~E., and Silander, J.~A. (2023).
\newblock Zero-inflated beta distribution regression modeling.
\newblock {\em Journal of Agricultural, Biological and Environmental Statistics}, 28(1):117--137.

\bibitem[Turnbaugh et~al., 2007]{biblio6}
Turnbaugh, P.~J., Ley, R.~E., Hamady, M., Fraser-Liggett, C.~M., Knight, R., and Gordon, J.~I. (2007).
\newblock The human microbiome project.
\newblock {\em Nature}, 449(7164):804--810.

\bibitem[Tyler et~al., 2014]{biblio7}
Tyler, A.~D., Smith, M.~I., and Silverberg, M.~S. (2014).
\newblock Analyzing the human microbiome: a “how to” guide for physicians.
\newblock {\em {A}merican {C}ollege of {G}astroenterology}, 109(7):983--993.

\bibitem[Wei and Tanner, 1990]{biblio51}
Wei, G. C.~G. and Tanner, M.~A. (1990).
\newblock A {Monte Carlo} implementation of the {EM} algorithm and the poor man's data augmentation algorithms.
\newblock {\em Journal of the American Statistical Association}, 85(411):699--704.

\bibitem[Zhang et~al., 2020]{biblio16}
Zhang, X., Guo, B., and Yi, N. (2020).
\newblock Zero-inflated gaussian mixed models for analyzing longitudinal microbiome data.
\newblock {\em PloS {ONE}}, 15(11):e0242073.

\bibitem[Zhang et~al., 2018]{biblio38}
Zhang, X., Pei, Y.-F., Zhang, L., Guo, B., Pendegraft, A.~H., Zhuang, W., and Yi, N. (2018).
\newblock Negative binomial mixed models for analyzing longitudinal microbiome data.
\newblock {\em Frontiers in Microbiology}, 9:1683.

\bibitem[{Zhang Chen}, 2023]{biblio27}
{Zhang Chen}, E. (2023).
\newblock {\em ZIBR: A Zero-Inflated Beta Random Effect Model}.
\newblock R package version 1.0.2.

\bibitem[Zhu and Lee, 2002]{biblio42}
Zhu, H.-T. and Lee, S.-Y. (2002).
\newblock Analysis of generalized linear mixed models via a stochastic approximation algorithm with {Markov chain Monte-Carlo} method.
\newblock {\em Statistics and Computing}, 12(2):175--183.

\end{thebibliography}

\end{document}


\maketitle
\vspace*{3cm}

\begin{appendices}

\section{Additional comparative simulation study on unbalanced and interpolated datasets}\label{secap1}

In this section we report some results obtained in the application of the ZIBR model to an unbalanced data scenario. In this context, some studies \citep{biblio33} advise the use of imputation with the individual average or to remove some observations in the data for the use of longitudinal data methods that cannot cope with unbalanced designs. Therefore, this is the approach we consider here to be able to compare GHQ to SAEM on this setting. However, as already mentioned, interpolation can create some inaccuracies in the estimations. In comparison, estimation using the SAEM algorithm does allow for its use directly on the original unbalanced data. We also compare the estimation obtained with the GAMLSS implementation, which does allow to work with unbalanced data.

\subsection{Setup}

In the context of an unbalanced data scenario, data generation will have two parts. First, we will use the parameters of Setting 2 ($a=b=-0.5$, $\alpha=\beta=0.5$, $\sigma_1=0.7$, $\sigma_2=0.5$, $\phi=6.4$.) to generate 1000 datasets, with $T_i=10$ and different values for the number of individuals, $N\in\{50,100,200\}$. The variables $X$ and $Z$ are defined as usual. Then, we will randomly eliminate 20\% of the observations from each data set; therefore, the median (interquartile range (IQR)) of the number of observations per individual in the three specifications is 8 (IQR: 7 to 9). Given the drop-out method we chose to simulate an unbalanced data situation, we can assume that we are in a case of MCAR (Missing Completely At Random) \citep{biblio44}. Finally, we will compare the performance of SAEM on unbalanced data with GHQ on interpolated data. For each individual, the interpolation process will be carried out as follows:

\begin{itemize}
    \item if the missing value is between two known observations, linear interpolation will be performed; and
    \item if the missing value is at the end of the observations, it will be replaced with the last known value.
\end{itemize}

As in the main article, we compute the bias, MAE and RMSE of the estimations with all methods and the estimated densities of the parameter estimates.

\subsection{Results}

\begin{figure*}[htbp]%
\centering
\includegraphics[width=0.9\textwidth]{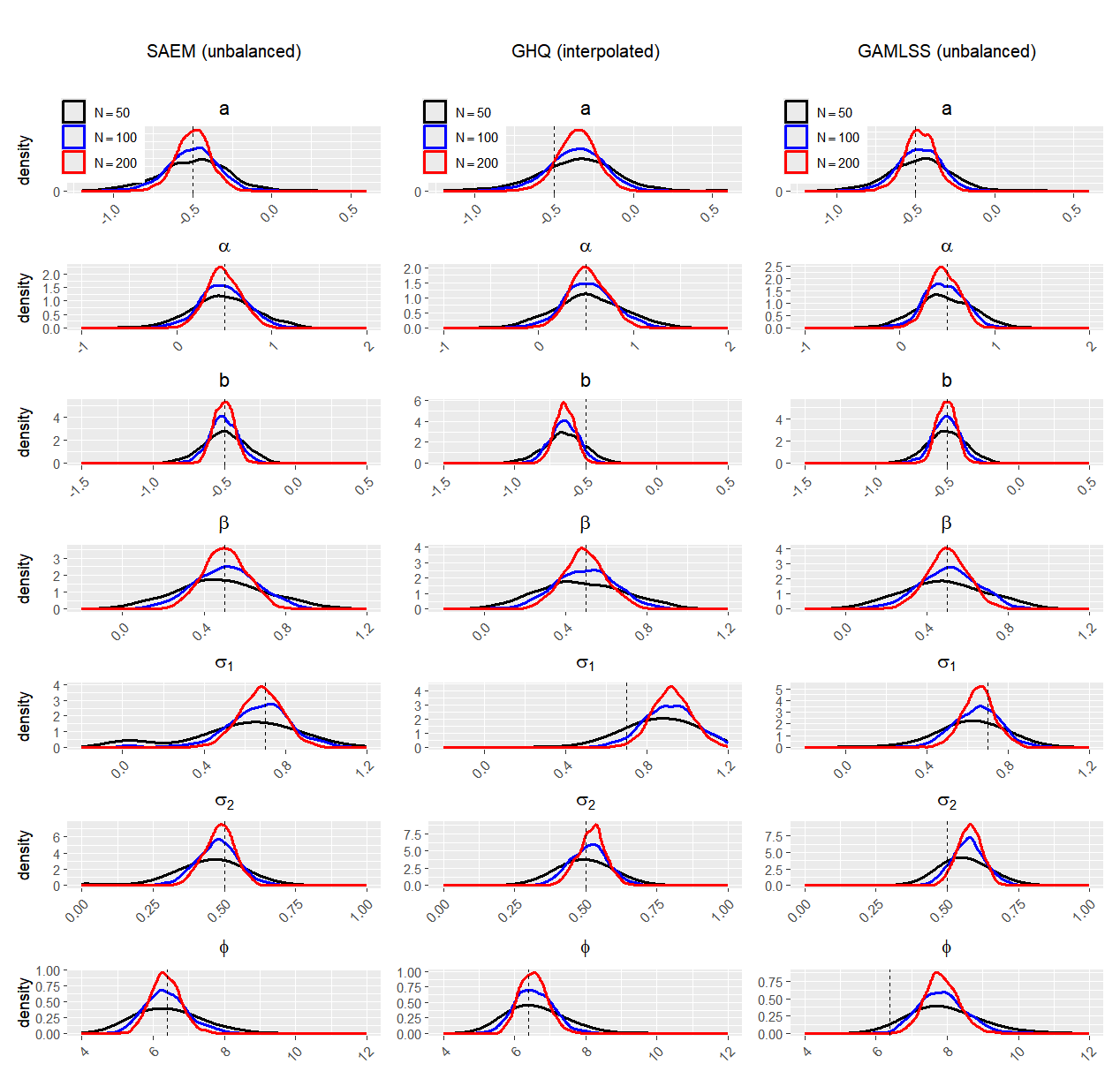}
\caption{Estimated density of the parameters obtained by the SAEM algorithm on unbalanced datasets and the GHQ procedure on interpolated datasets simulated under Setting 2. The dotted vertical line represents the true value of the parameter.}\label{apfig1}
\end{figure*}

First of all, notice that, from the two simulations scenarios used in the analysis of balanced data (Section 3 of the main document), the one used here (Setting 2) was the most favorable to the GHQ method. Table \ref{aptab1} shows the results of the estimation on unbalanced datasets for the SAEM algorithm and the GAMLSS method, and on interpolated datasets with the GHQ procedure. These results show that in most cases, the SAEM estimators outperform the GHQ and GAMLSS estimators by having a lower absolute bias. Furthermore, in the case of both SAEM and GAMLSS, the increase in the number of individuals reduces the MAE and RMSE values in all cases; something that cannot be said for the estimates obtained by GHQ. Although there are some scenarios in which GHQ obtains better results in error measures, this advantage is quite small, whereas when SAEM is superior, the differences in the values are much more noticeable, which can be clearly seen in Figure \ref{apfig1}. In this scenario, the estimate using GAMLSS presents the best RMSE and MAE values of the three methods in most cases, although this advantage is generally quite small. Additionally, as the sample size increases, these estimates gradually approach those obtained by SAEM.

\begin{table}[htbp]
\footnotesize
\caption{Summary statistics of the results obtained by the SAEM algorithm and GAMLSS method on unbalanced and GHQ procedure on interpolated data sets over 1000 simulation runs. For each parameter value and number of individuals, $N$, bold numbers indicate the lowest (absolute) value for each of bias, RMSE and MAE.}\label{aptab1}
\begin{tabular*}{\textwidth}{@{\extracolsep\fill}lclccccccccc}
\hline
\textbf{Parameter} & \textbf{Value} & & \textbf{Bias} & \textbf{RMSE} & \textbf{MAE} & \textbf{Bias} & \textbf{RMSE} & \textbf{MAE}& \textbf{Bias} & \textbf{RMSE} & \textbf{MAE}\\
\toprule

&&&\multicolumn{3}{c}{\textbf{SAEM (unbalanced)}}&\multicolumn{3}{c}{\textbf{GHQ (interpolated)}}&\multicolumn{3}{c}{\textbf{GAMLSS (unbalanced)}}\\
\cmidrule{4-6}\cmidrule{7-9}\cmidrule{10-12}

$a$ & -0.5 & $N=50$ & {\bf 0.0048} & 0.2083 & 0.1662 & 0.1425 & 0.2726 & 0.2226 & 0.0264 & {\bf 0.1874} & {\bf 0.1494}\\
 & & $N=100$ & {\bf 0.0015} & 0.1508 & 0.1199 & 0.1408 & 0.2206 & 0.1826 & 0.0275 & {\bf 0.1338} & {\bf 0.1081}\\
 & & $N=200$ & {\bf 0.0041} & 0.1041 & 0.0830 & 0.1459 & 0.1869 & 0.1589 & 0.0283 & {\bf 0.0971} & {\bf 0.0774}\vspace*{1mm}\\

$\alpha$ & 0.5 & $N=50$ & {\bf -0.0107} & 0.3419 & 0.2694 & 0.0329 & 0.3782 & 0.2965 & -0.0279 & {\bf 0.3075} & {\bf 0.2445}\\
 & & $N=100$ & {\bf -0.0052} & 0.2450 & 0.1949 & 0.0358 & 0.2680 & 0.2109 & -0.0296 & {\bf 0.2125} & {\bf 0.1714}\\
 & & $N=200$ & {\bf -0.0069} & 0.1801 & 0.1434 & 0.0296 & 0.2002 & 0.1584 & -0.0314 & {\bf 0.1610} & {\bf 0.1299}\vspace*{1mm}\\

$b$ & -0.5 & $N=50$ & {\bf -0.0016} & 0.1449 & 0.1156 & -0.1495 & 0.2052 & 0.1689 & -0.0028 & {\bf 0.1340} & {\bf 0.1073}\\
 & & $N=100$ & {\bf -0.0047} & 0.0977 & 0.0777 & -0.1489 & 0.1765 & 0.1532 & -0.0048 & {\bf 0.0928} & {\bf 0.0736}\\
 & & $N=200$ & {\bf -0.0031} & 0.0685 & 0.0555 & -0.1476 & 0.1621 & 0.1481 & -0.0045 & {\bf 0.0654} & {\bf 0.0531}\vspace*{1mm}\\

$\beta$ & 0.5 & $N=50$ & -0.0118 & 0.2283 & 0.1828 & -0.0163 & 0.2171 & 0.1766 & {\bf -0.0096} & {\bf 0.2131} & {\bf 0.1714}\\
 & & $N=100$ & 0.0073 & 0.1578 & 0.1257 & {\bf -0.0015} & 0.1497 & 0.1203 & 0.0070 & {\bf 0.1484} & {\bf 0.1177}\\
 & & $N=200$ & 0.0042 & 0.1084 & 0.0864 & -0.0054 & 0.1064 & 0.0842 & {\bf 0.0022} & {\bf 0.1012} & {\bf 0.0801}\vspace*{1mm}\\

$\sigma_1$ & 0.7 & $N=50$ & -0.1083 & 0.2935 & 0.2164 & 0.1867 & 0.2706 & 0.2205 & {\bf -0.0787} & {\bf 0.1859} & {\bf 0.1443}\\
 & & $N=100$ & {\bf -0.0288} & 0.1637 & 0.1226 & 0.2223 & 0.2606 & 0.2281 & -0.0507 & {\bf 0.1246} & {\bf 0.0984}\\
 & & $N=200$ & {\bf -0.0180} & 0.1063 & 0.0841 & 0.2248 & 0.2441 & 0.2253 & -0.0447 & {\bf 0.0912} & {\bf 0.0723}\vspace*{1mm}\\

$\sigma_2$ & 0.5 & $N=50$ & -0.0532 & 0.1411 & 0.1034 & {\bf -0.0038} & {\bf 0.0960} & {\bf 0.0764} & 0.0626 & 0.1059 & 0.0838\\
 & & $N=100$ & -0.0245 & 0.0797 & 0.0616 & {\bf 0.0120} & {\bf 0.0685} & {\bf 0.0546} & 0.0759 & 0.0961 & 0.0811\\
 & & $N=200$ & {\bf -0.0173} & 0.0554 & 0.0438 & 0.0210 & {\bf 0.0526} & {\bf 0.0425} & 0.0792 & 0.0897 & 0.0799\vspace*{1mm}\\

$\phi$ & 6.4 & $N=50$ & {\bf 0.0405} & 0.9219 & 0.7313 & 0.2200 & {\bf 0.8603} & {\bf 0.6580} & 1.5476 & 1.8430 & 1.5695\\
 & & $N=100$ & {\bf -0.0186} & 0.6004 & 0.4772 & 0.1560 & {\bf 0.5562} & {\bf 0.4419} & 1.4776 & 1.6241 & 1.4802\\
 & & $N=200$ & {\bf -0.0454} & 0.4161 & 0.3334 & 0.1330 & {\bf 0.4030} & {\bf 0.3204} & 1.4547 & 1.5270 & 1.4547\vspace*{1mm}\\
\bottomrule

\end{tabular*}
\end{table}
\vspace*{5mm}

It is interesting to note that the worst results of GHQ are concentrated in the parameters related to the logistic part of the ZIBR model; that is, the one that controls zero inflation. This could be evidence that interpolation affects this component of the model much more than the other. In Figure \ref{apfig1}, where the densities of the estimators obtained by the methods are shown, we see a behavior that confirms what was mentioned, being also clear the fact that the random components $a$ and $b$ are those that show a distribution much further from the theoretical values for the GHQ method with interpolation. Also the variance components show greater deviation from the real value according to the density graph. For GAMLSS, its densities are comparable to those of SAEM in unbalanced datasets, with the exception of the $\sigma_2$ and $\phi$ parameters, which are specifically responsible for regulating the dispersion in the beta part of the ZIBR model.

Lastly, although a comparison with the results obtained on balanced data sets simulated under Setting 2 can not be directly established, since the number of observations differ, notice that the estimation results obtained with SAEM in the two cases (Figure \ref{apfig1} here and Figure 2 in the main document) are quite similar, whereas for GHQ there is a clear impact of interpolation on the estimation results.

\section{Additional simulation study of hypothesis tests on covariates association}\label{secap2}

In the context of the ZIBR model, the effect of covariates on the presence or absence and abundance of a bacteria taxon must be studied. Therefore, we need procedures to test the null hypothesis $H_0:\alpha=\beta=0$, $H_0:\alpha=0$ and $H_0:\beta=0$. A common approach in mixed models is to use the Likelihood Ratio Test (LRT) in this context, as a way of comparing two nested models. An alternative way to test both fixed and random effects parameters significance is to use the Wald test, for which standard errors estimates of each parameter are required.

\subsection{Likelihood ratio test}\label{secap21}

The likelihood ratio test is performed to test the null hypothesis $H_0:\alpha=\beta=0$. We will now analyze its type I error with the SAEM estimation method. As for parameter estimation, we are interested in the performance on both balanced and unbalanced data, and we will compare it with the results of the LRT based on the GHQ procedure, only in the balanced data scenario. However, as can be seen in \cite{biblio56}, the calculation of the log-likelihood of the ZIBR model using GAMLSS is not the same as in the other two procedures, since it uses estimators based on a penalized quasi-likelihood method \citep{biblio58}. For this reason, the values obtained for the log-likelihood in GAMLSS are not comparable with the other methods and the Likelihood Ratio Test cannot be applied in the same way. Therefore, we will not report it in this section. 

The parameter values to simulate 1000 datasets are set as follows:

\begin{itemize}
    \item $a=-0.5, b=0.5$,
    \item $\alpha=\beta=0$,
    \item $\sigma_1=0.7$, $\sigma_2=0.5$,
    \item $\phi=6.4$.    
\end{itemize}

The number of individuals in each dataset takes two values $N\in\{50,100\}$, and we keep the number of observations per individual fixed $T_i=10$. The procedure will be carried out on both 1000 balanced datasets and 1000 datasets from which 20\% of their observations are dropped out following the MCAR process already described. The SAEM estimation settings were the same as above, while the log-likelihood estimation by Importance Sampling was carried out by simulating $K=500$ values.

The results shown in Table \ref{aptab2} make evident the similar behavior in balanced and unbalanced data with regard to the LRT procedure. In no case is there a significant difference between the expected and obtained values, and this behavior is not affected by the number of individuals or the type of data being worked on. Both SAEM and GHQ obtain good results in the balanced case, while in the unbalanced case SAEM approximates the theoretical level quite well.

\begin{table}[htb]
\centering
\caption{Type I error for testing $H_0:\alpha=\beta=0$ in balanced and unbalanced data with the SAEM algorithm and the GHQ procedure for nominal significance level of 0.05 and 0.01.}\label{aptab2}
\begin{tabular*}{0.6\textwidth}{@{\extracolsep\fill}llcccc}
\hline
&&\multicolumn{2}{@{}c@{}}{\textbf{SAEM}}&\multicolumn{2}{@{}c@{}}{\textbf{GHQ}} \\
\cmidrule{3-6}
&&\multicolumn{2}{@{}c@{}}{\textbf{Significance level}}&\multicolumn{2}{@{}c@{}}{\textbf{Significance level}}\\
Data type && $0.05$ & $0.01$ & $0.05$ & $0.01$\\
\midrule
Balanced & $N=50$ & 0.050 & 0.007 & 0.059 & 0.009\\
 & $N=100$ & 0.048 & 0.009 & 0.050 & 0.012\\
Unbalanced & $N=50$ & 0.064 & 0.009 &&\\
 & $N=100$ & 0.045 & 0.010 &&\\
\bottomrule
\end{tabular*}
\vspace*{5mm}
\end{table}

\subsection{Wald test for individual parameters}\label{secap22}

Finally, we will check the results of the Wald test  using the standard errors of the estimators computed by stochastic approximation of the Fisher observed matrix. It is important to mention that the GHQ method implemented in the \texttt{ZIBR} package does not offer any method for calculating the standard errors of the estimates, so the results obtained using SAEM cannot be compared with those provided by the GHQ method. However, GAMLSS does produce estimates for the standard errors of the estimators, so comparisons can be made between SAEM and GAMLSS.

\subsubsection{Fixed effects}\label{secap221} 

The null hypotheses to be tested are $H_0: \alpha=0$ and $H_0: \beta=0$, for which the test statistics are defined by
$$\frac{\hat{\alpha}^2}{Var(\hat{\alpha})}\ \ \ \mbox{ and }\ \ \  \frac{\hat{\beta}^2}{Var(\hat{\beta})}$$
where $Var(\hat{\alpha})$ and $Var(\hat{\beta})$ are estimated by the procedure described in Section 2.3 of the main document. Under the null hypothesis, these variables follow an asymptotic $\chi^2$ distribution with one degree of freedom.

We keep the simulation settings used for the LRT. To improve the convergence properties of the SAEM algorithm, we use $m=10$ Markov chains in the execution. The results are summarized in Table \ref{aptab3}. We can see that the values obtained through simulation are not far from the theoretically expected values, although slightly above the ones obtained with the Likelihood Ratio Test. As shown in the table, SAEM achieves type I errors close to the theoretical values of the test, while GAMLSS does not achieve optimal results. Furthermore, the inaccuracies of GAMLSS are more pronounced for testing $\beta=0$ than testing $\alpha=0$, which confirms the tendency of GAMLSS to produce poor results in the Beta component of the ZIBR model.

\begin{table}[htb]
\centering
\caption{Type I error of the Wald test for $H_0:\alpha=0$ and $H_0:\beta=0$ using the SAEM algorithm for nominal significance level of 0.05 and 0.01.\vspace*{2mm}}\label{aptab3}
\begin{tabular*}{0.6\textwidth}{@{\extracolsep\fill}lcccc}
\hline
&\multicolumn{2}{@{}c@{}}{\textbf{SAEM}}&\multicolumn{2}{@{}c@{}}{\textbf{GAMLSS}} \\
\cmidrule{2-5}
&\multicolumn{2}{@{}c@{}}{\textbf{Significance level}}&\multicolumn{2}{@{}c@{}}{\textbf{Significance level}}\\
 & $0.05$ & $0.01$& $0.05$ & $0.01$\\
\midrule
$H_0: \alpha=0$&0.069&0.022&0.120&0.038\\
$H_0: \beta=0$&0.062&0.013&0.296&0.158\\
\bottomrule
\end{tabular*}
\end{table}

\subsubsection{Random effects}\label{secap222}

The null hypotheses are now $H_0: a=0$ and $H_0: b=0$, and the corresponding test statistics
$$\frac{\hat{a}^2}{Var(\hat{a})}\ \ \ \mbox{ and }\ \ \ \frac{\hat{b}^2}{Var(\hat{b})},$$
follow each a $\chi^2$ distribution with one degree of freedom, if the null hypothesis are valid. The simulation settings now are those of Setting 2 with $a=0$ and $b=0$. As for the fixed-effects test, we use $m=10$ Markov chains to accelerate convergence. The results of the test are presented in Table \ref{aptab4}.

\begin{table}[htb]
\centering
\caption{Type I error of the Wald test for $H_0:a=0$ and $H_0:b=0$ using the SAEM algorithm for nominal $\alpha$-level of 0.05 and 0.01.\vspace*{2mm}}\label{aptab4}
\begin{tabular*}{0.6\textwidth}{@{\extracolsep\fill}lcccc}
\hline
&\multicolumn{2}{c}{\textbf{SAEM}}&\multicolumn{2}{c}{\textbf{GAMLSS}} \\
\cmidrule{2-5}
&\multicolumn{2}{@{}c@{}}{\textbf{Significance level}}&\multicolumn{2}{@{}c@{}}{\textbf{Significance level}}\\
 & $0.05$ & $0.01$& $0.05$ & $0.01$\\
\midrule
$H_0: a=0$&0.051&0.014&0.092&0.029\\
$H_0: b=0$&0.049&0.011&0.250&0.126\\
\bottomrule
\end{tabular*}
\vspace*{5mm}
\end{table}

The results of this section, in the same way as those of the previous one, are indicative that the calculation of the standard errors given by SAEM meets the expected statistical properties, while GAMLSS does not obtain similar Type I errors. One of the advantages of this result is that it allows the development of hypothesis tests in a less computationally demanding way than the Likelihood Ratio Test, since this requires the estimation of two different models, while the Wald test only needs to estimate one.

\section{Additional simulation study to evaluate the power of SAEM and GHQ as a function of the proportion of zeros}
\label{secapZI}

Following \cite{biblio27}, we conducted a series of simulations to evaluate how the statistical power of the SAEM and GHQ methods varies with the proportion of zeros in the data, which is controlled by the parameter $\alpha_0$. The intercept $\alpha_0$ determines the fraction of zeros generated in the simulated datasets. Four parameter configurations for $(\alpha_0, \alpha_1, \beta_0, \beta_1)$ were considered:
\begin{itemize}
\item Setting 1: $([-1, 1], 0.5, 0.5, 0.5)$,
\item Setting 2: $([-1, 1], -0.5, 0.5, 0.5)$,
\item Setting 3: $([-1, 1], 0, 0.5, 0.5)$,
\item Setting 4: $([-1, 1], 0.5, 0, 0)$.
\end{itemize}
Following the simulation design described in Appendix~\ref{secap1}, we simulated $N = 50$ individuals with $T = 10$ observations per individual. The variances of the mixed-effects components were set to $\sigma_1 = 0.7$ and $\sigma_2 = 0.5$, and the dispersion parameter of the Beta distribution was fixed at $6.4$. As described in Section~3, a covariate $X$ was defined to represent a treatment–control structure, assigning $X = 0$ to the first half of the individuals and $X = 1$ to the remaining half. The same variable was used as a covariate in both components of the model, setting $Z = X$. The simuation for each $\alpha_0$ value was repeated 1 000 times, and for each setting, we tested the null hypothesis $H_0 : \alpha_1=\beta_1=0$ using an $\alpha$-level of 0.05.

\begin{figure}[htbp]%
\centering
\includegraphics[width=0.4\textwidth]{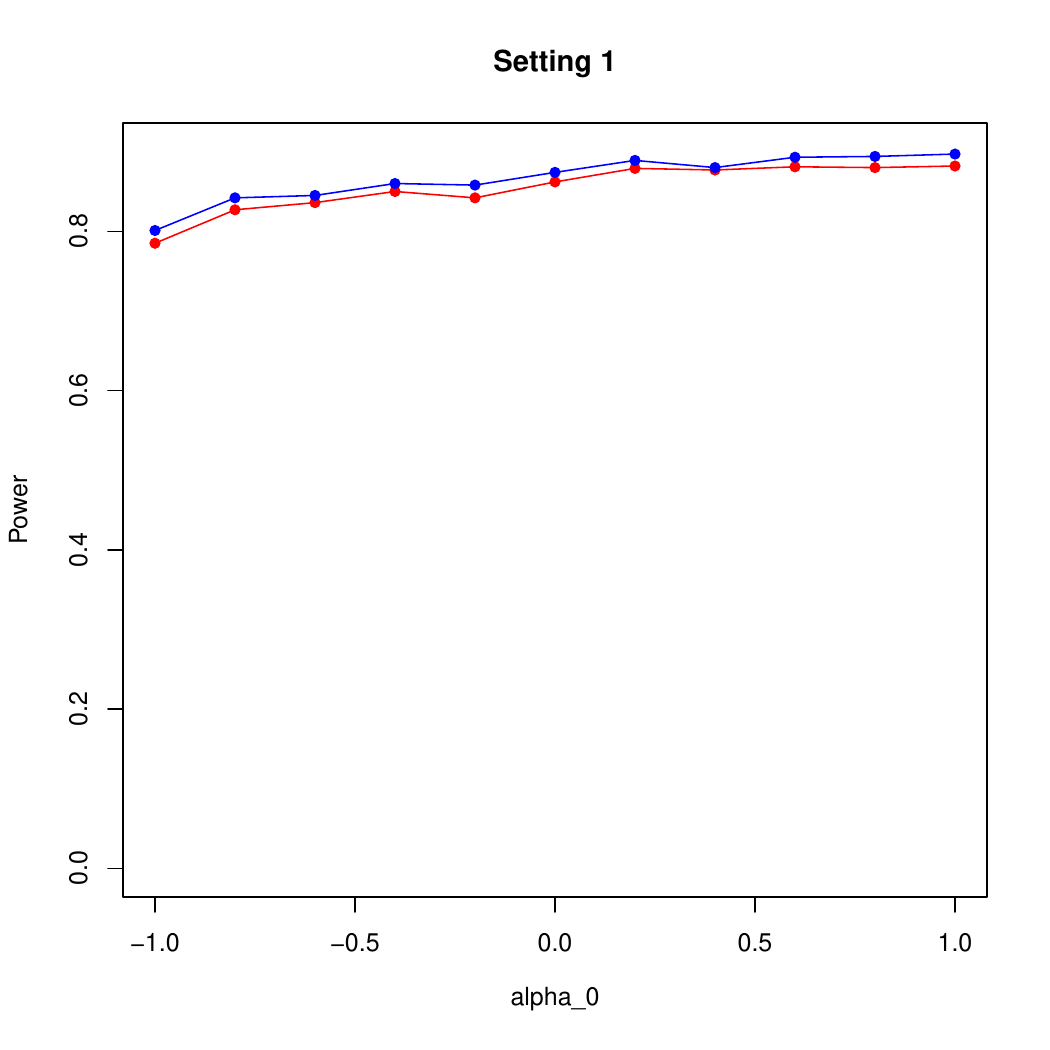}
\includegraphics[width=0.4\textwidth]{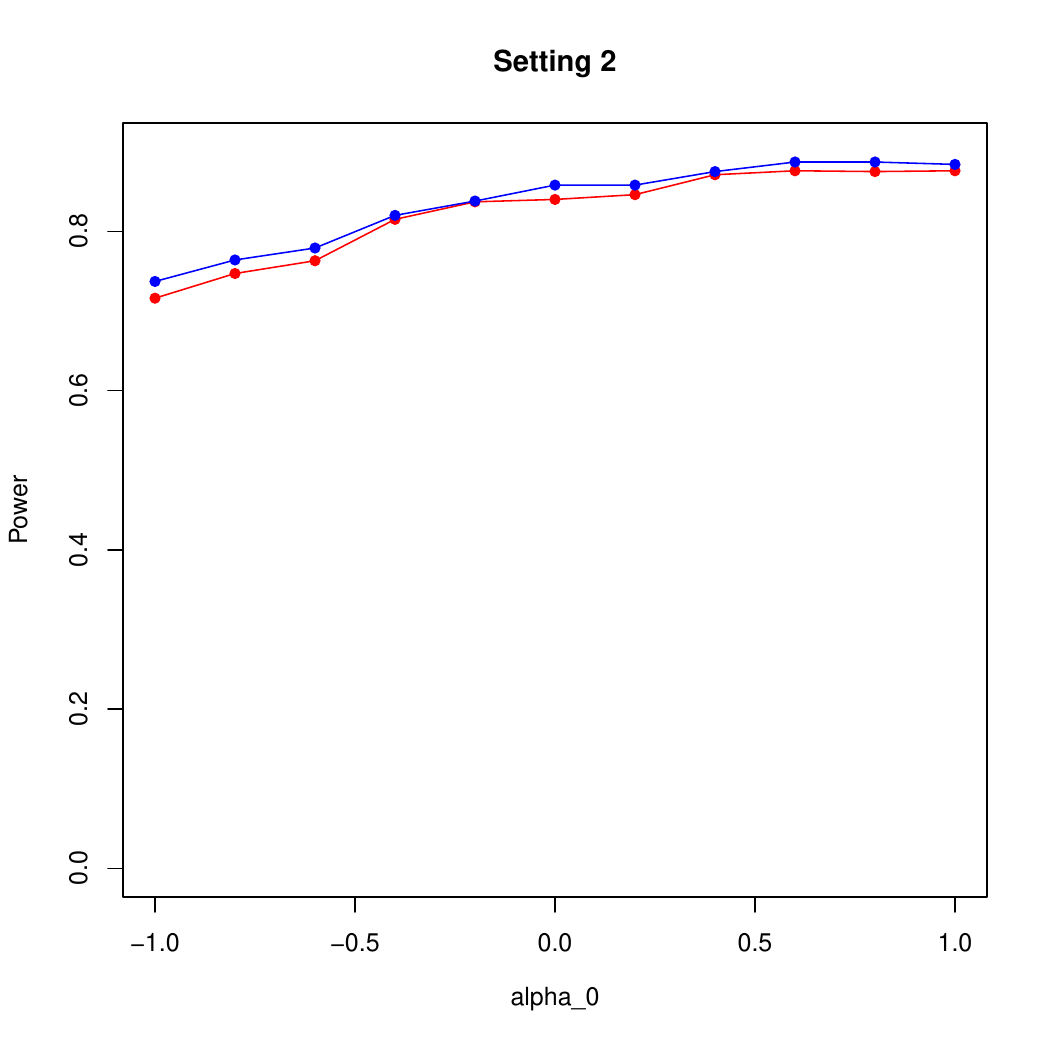}
\includegraphics[width=0.4\textwidth]{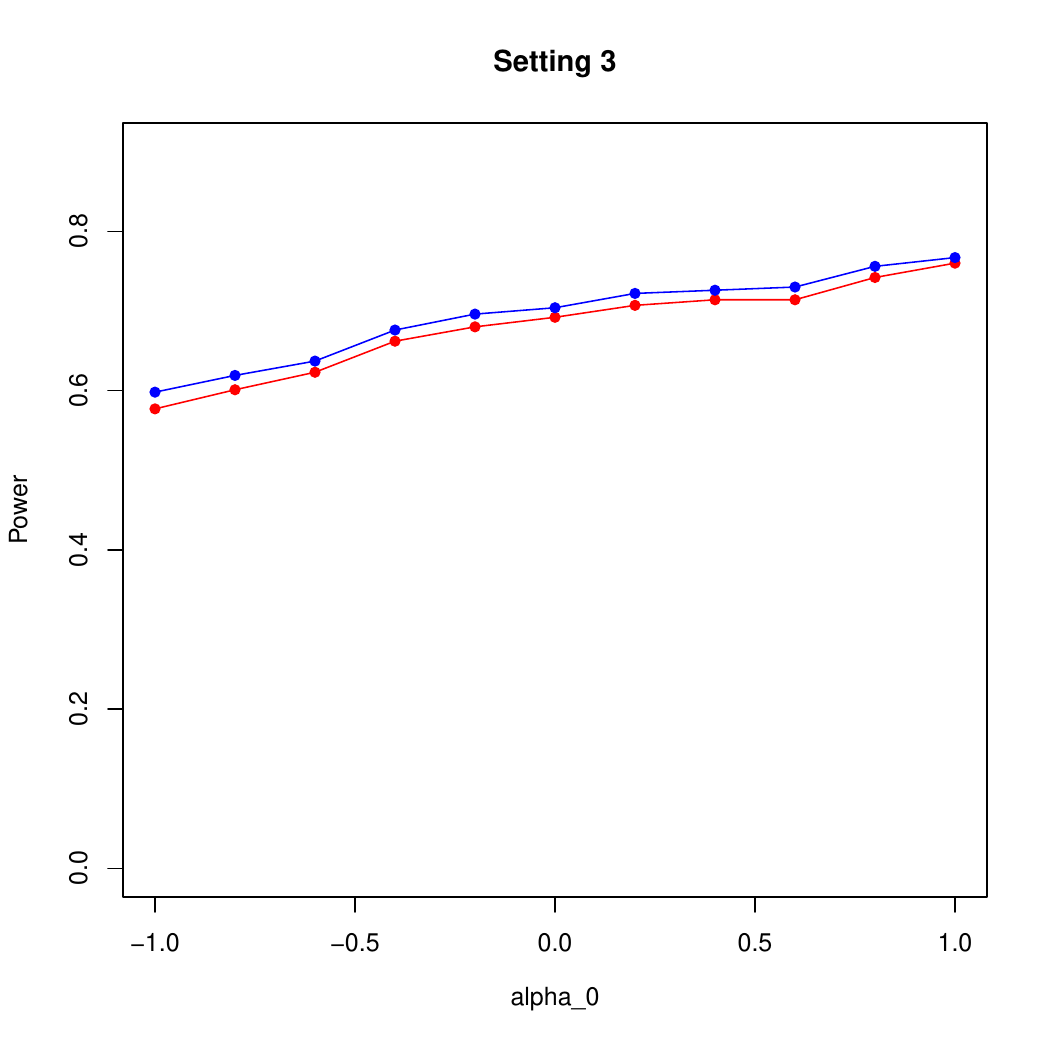}
\includegraphics[width=0.4\textwidth]{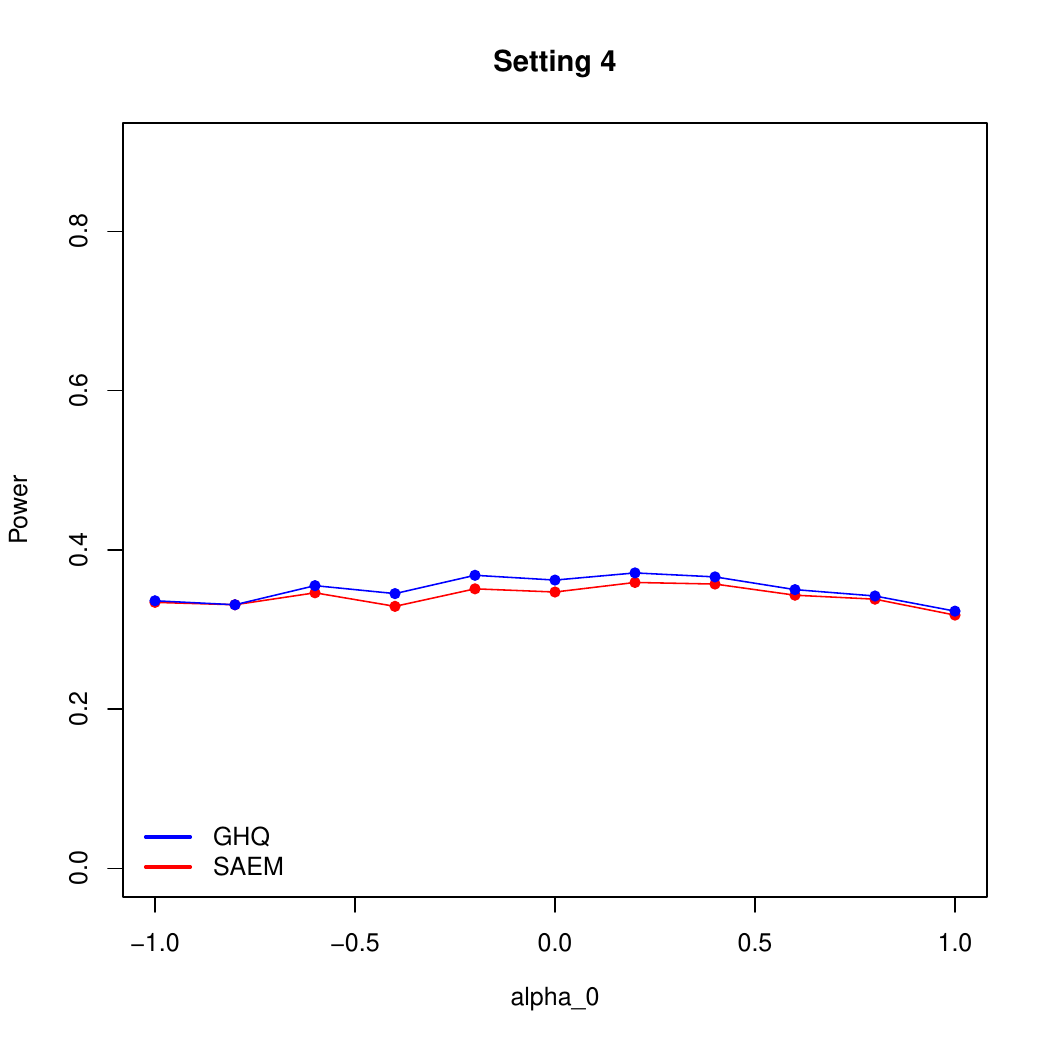}
\caption{Proportion of presence of the taxa in the observations of the two groups of women (pregnant and non-pregnant) and the difference between these values}\label{power}
\end{figure}

The parameter $\alpha_0$ was systematically varied within the specified range to construct the power curves. Figure~\ref{power} displays these curves as functions of $\alpha_0$, showing that increasing $\alpha_0$ corresponds to a decreasing proportion of zeros in the data.

In Figure \ref{power}, the power curves obtained under the four simulated scenarios illustrate how both SAEM and GHQ respond to changes in the proportion of zeros governed by the intercept parameter $\alpha_0$. Both methods produce extremely similar results, replicating the same overall trend. Across settings, the statistical power generally increases as $\alpha_0$ rises (indicating fewer zeros) except for Setting 4. We can also observe that both methods achieve greater and more stable power, especially when there are associations between the logistic and continuous components (Settings 1 and 2). When we assumed no association in logistic component (Setting 3), lower power is observed. The worst performance is observed when no association is assumed in the Beta component (Setting 4).

\section{Aditional case study: Pregnancy effect in vaginal microbiome}\label{secap3}

In this case we apply the ZIBR model to the analysis of longitudinal data from a study \citep{biblio32} describing the vaginal microbiome of a group of 22 pregnant and 32 non-pregnant women. In this case we try to verify the effect of pregnancy on the distribution of the different bacterial taxa observed, in a similar way to what is done in a recent work \citep{biblio16}. However, unlike the cited study, where the analysis is performed on count data, we will analyse the data as proportions using the ZIBR model with SAEM estimation. Furthermore, a comparison of the results with the GHQ method can not be established since the number of time points is different between individuals; that is, the data is unbalanced.

\begin{table}[htbp]
\centering
\caption{Characteristics of the two groups of women, separated by pregnancy status}\label{aptab7}
\begin{threeparttable}
\begin{tabular*}{0.7\textwidth}{@{\extracolsep\fill}lcc}
\hline%
 & Non-pregnant & Pregnant\\
Variable$^{1}$&$N = 32$&$N = 22$\\
\midrule
Age (years) & 37 (31-43) & 24 (20-29) \\ 
Time (months) & 3.40 (3.52-3.67) & 8.13 (8.00-8.43) \\ 
N. of observations & 24 (21-29) & 6 (6-7) \\ 
\bottomrule
\end{tabular*}
\begin{tablenotes}%
\footnotesize
\item[$^{1}$] Mean(Q1-Q3)
\end{tablenotes}
\end{threeparttable}
\end{table}

A preliminary review of the data (Table \ref{aptab7}) allows us to notice that there are large differences between the characteristics of pregnant and non-pregnant women. The time span of observations is much longer for pregnant women, and thus they have many more readings than non-pregnant women. In addition, the average age of pregnant women is much younger than the non-pregnant group. This is why we decided to include age as a covariate in the models to be used, according to the following specifications:

\begin{itemize}
    \item \textbf{Model 1:} pregnancy, time and age as covariates, taking pregnancy as a factor of interest for testing.
    \item \textbf{Model 2:} pregnancy, time, age and interaction between time and pregnancy as covariates, testing the effect of pregnancy and the interaction.
\end{itemize}

Once we filter the bacterial taxa with a proportion of zeros between 0.1 and 0.9 and those that are absent in either of the two groups of women, we have 897 observations from 54 individuals and 57 taxa. With these data, we developed the LRT for the proposed variables in each model using SAEM at a significance level of 0.05. 

\begin{figure}[htbp]%
\centering
\includegraphics[width=0.9\textwidth]{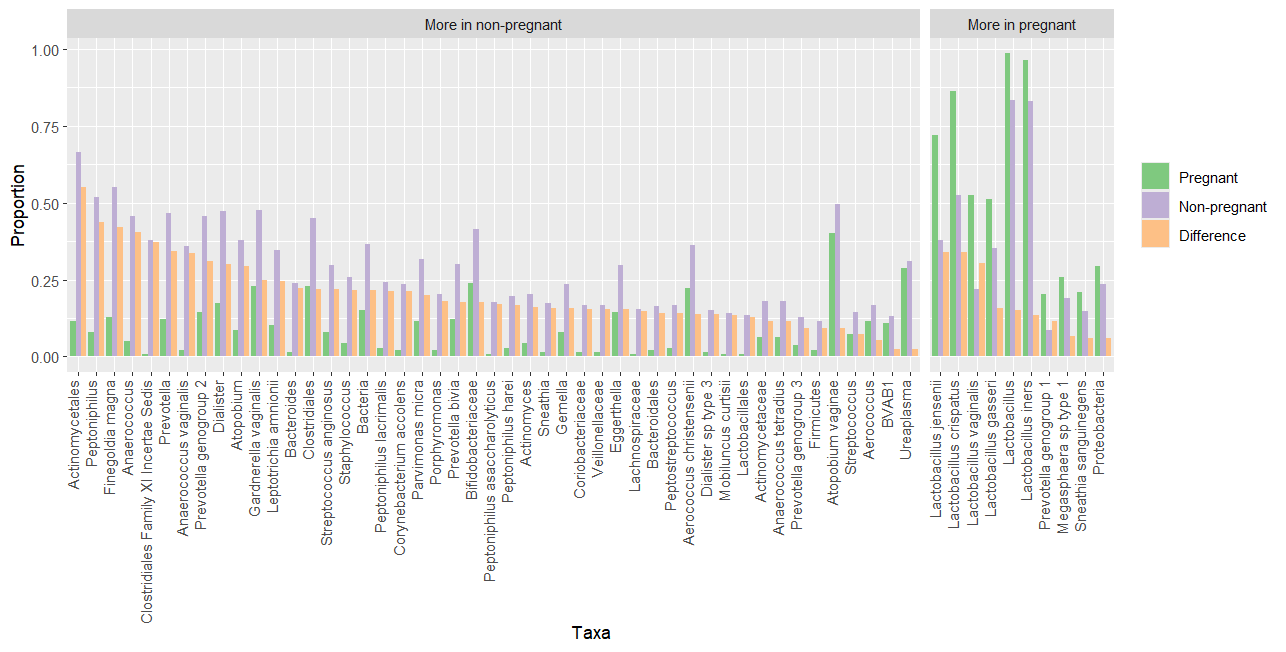}
\caption{Proportion of presence of the taxa in the observations of the two groups of women (pregnant and non-pregnant) and the difference between these values}\label{apfig4}
\end{figure}

\begin{figure}[htbp]%
\centering
\includegraphics[width=0.9\textwidth]{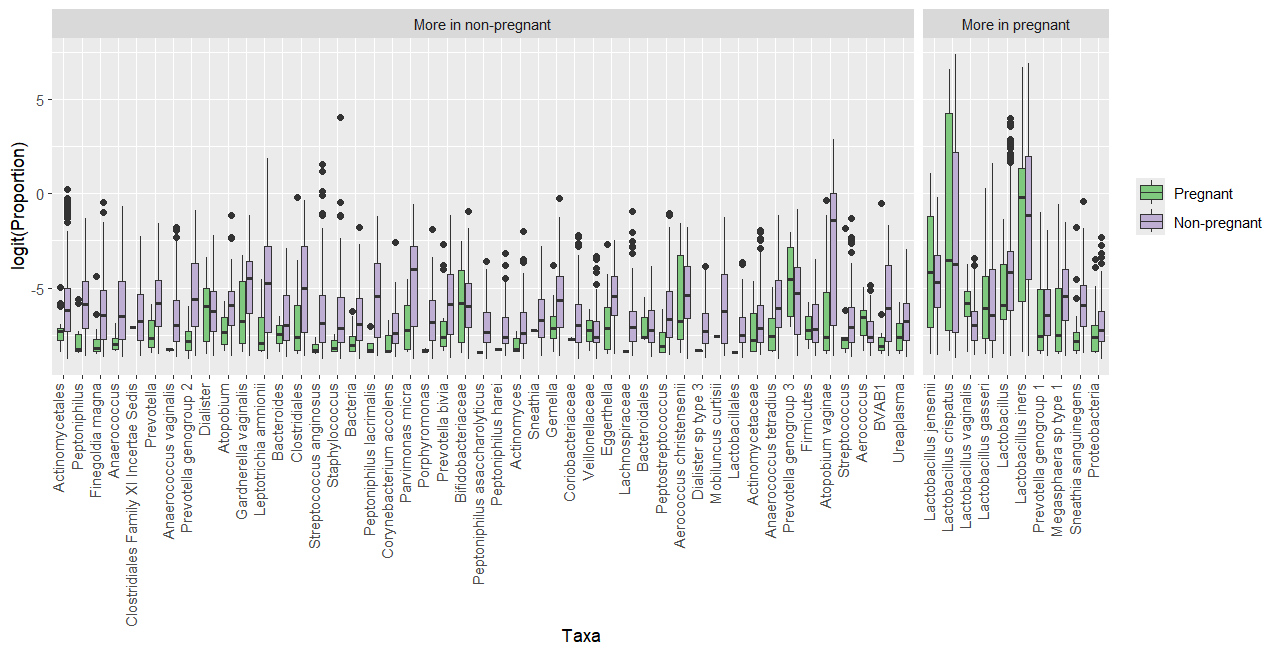}
\caption{Logit of the non-zero abundance of the taxa in the observations of the two groups of women (pregnant and non-pregnant)}\label{apfig5}
\end{figure}

Figures \ref{apfig4} and \ref{apfig5} show the differences in presence of the taxa considered and the distribution of the non-zero abundance data. From these figures we can see that, on average, there are more bacteria with a higher abundance in non-pregnant women. In pregnant women, however, the dominance of the \textit{Lactobacillus} genus in the most abundant bacteria is very evident, which is consistent with previous findings \citep{biblio45}, which also suggest that low variety and high stability is another characteristic of the vaginal microbiome in pregnant women. 

\begin{figure}[htbp]%
\centering
\includegraphics[width=\textwidth]{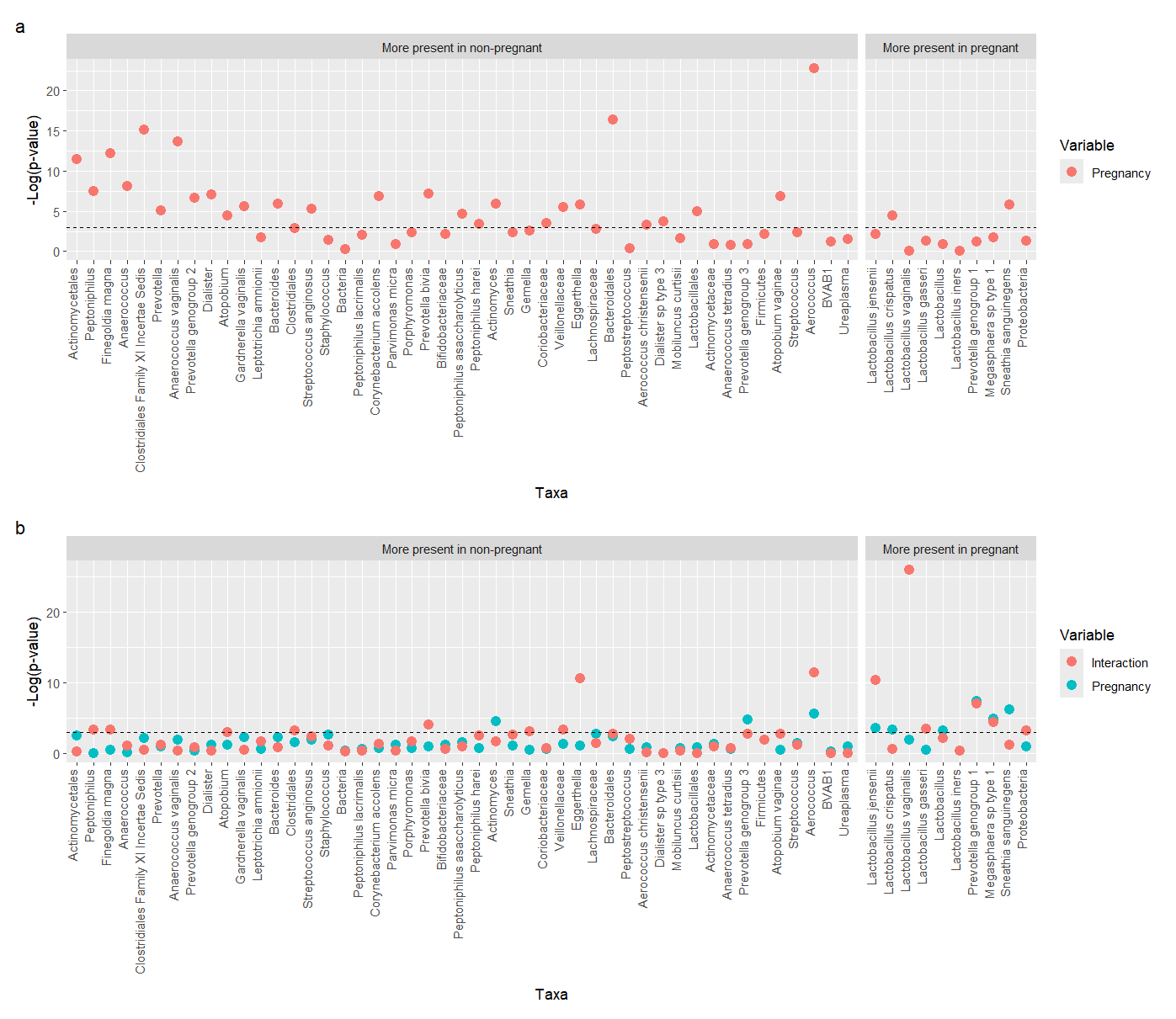}
\caption{Negative of log transformed p-value of the LRT for the interest variables in Model 1 (a) and Model 2 (b) for the bacterial taxa. The horizontal line represents the threshold $\alpha=0.05$}\label{fig5}
\end{figure}

The LRT indicates that Model 1 was able to detect a higher number of bacteria (51\% of all taxa) affected by pregnancy than Model 2 (16\%), for which the interaction of time with pregnancy is statistically significant for a higher number of bacteria (26\%) compared to pregnancy alone. Figure \ref{fig5} details these results further. In particular, for Model 1 the bacteria for which an influence of pregnancy is detected are more commonly found among those that are more abundant in non-pregnant women. In comparison, only 2 of the most present bacteria in pregnant women show a statistical significance of pregnancy: \textit{Lactobacillus crispatus} and \textit{Sneathia sanguinegens}. The information in Table \ref{aptab8} also confirms these findings, showing that the coefficients associated with pregnancy for these taxa in the abundance part have different sign. In a previous work \citep{biblio32} it is found that bacteria of the genus \textit{Sneathia}, potentially pathogenic, reduce their presence during pregnancy, while \textit{Lactobacillus crispatus} abundance in pregnancy is associated with a lower risk of preterm delivery \citep{biblio46}. 

However, Figure \ref{fig5}b shows that adding the time-pregnancy interaction to the model specification changes the results. First, only a small number of the bacteria that predominate in non-pregnant women show significant dependence on pregnancy or the interaction between time and pregnancy. But in the other group of bacteria, nine show significance for pregnancy or the interaction, and three of them for both: \textit{Lactobacillus jensenii}, \textit{Prevotella genogroup 1} and \textit{Megasphaera sp type 1}. Previous studies \citep{biblio47} report that both \textit{Lactobacillus} bacteria and those associated with bacterial vaginosis (\textit{Prevotella}, \textit{Sneathia}) change their abundance between pregnant and non-pregnant women and also along time in case of pregnancy. The coefficients associated with the variables can be consulted in Table \ref{aptab9}. In view of these results, we can assert that the ZIBR model and the SAEM estimation for relative abundance data obtain similar conclusions as both previous research results and the mixed models defined for log-transformed count data.

\begin{table}[htbp]
\footnotesize
\centering
\caption{ML estimates calculated by the SAEM algorithm for the parameters of Model 1 on the vaginal microbiome data}\label{aptab8}

\begin{threeparttable}
\begin{tabular}{rcccccc}
\hline
&\multicolumn{3}{@{}c@{}}{\textbf{Logistic part}}&\multicolumn{3}{@{}c@{}}{\textbf{Beta part}}\\
\cmidrule{2-4}\cmidrule{5-7}
\multicolumn{1}{c}{\textbf{Taxa}} & \textbf{Time} & \textbf{Pregnancy} & \textbf{Age}$^{1}$ & \textbf{Time} & \textbf{Pregnancy} & \textbf{Age}$^{1}$\\
\midrule
\multicolumn{1}{l}{\textbf{More present in non-pregnant}}&&&&&&\\
Actinomycetales & 0.1643 & \textbf{-2.9565} & -4.3481 & 0.2711 & \textbf{-0.5074} & 0.1593\\
Peptoniphilus &  -0.5538 & \textbf{-3.3886} & -4.9629 & 0.3549 & \textbf{-0.5902} & 0.4930\\
Finegoldia magna & 0.2397 & \textbf{-3.0966} & -4.3831 & -0.2211 & \textbf{-0.4315} & -0.0565\\
Anaerococcus &  0.3107 & \textbf{-3.1733} & -4.4771 & -0.0889 & \textbf{-0.4276} & 0.1836\\
Clostridiales Family XI Incertae Sedis  & 0.6038 & \textbf{-5.0301} & -5.4008 & -0.2489 & \textbf{-0.1186} & 0.4467\\
Prevotella & 0.0793 & \textbf{-2.7107} & -4.7978 & 0.0155 & \textbf{-0.5715} & 0.0004\\
Anaerococcus vaginalis & 0.8104 & \textbf{-4.6513} & -5.3721 & -0.2472 & \textbf{-0.3762} & 0.2780\\
Prevotella genogroup 2 & 0.5090 & \textbf{-1.6274} & -4.0081 & 0.7375 & \textbf{-1.2336} & -0.3287\\
Dialister & 2.2496 & \textbf{-2.9795} & -5.6544 & 0.4761 & \textbf{-0.2715} & 0.0820\\
Atopobium &  2.0765 & \textbf{-5.5589} & -5.2281 & -0.2565 & \textbf{-0.5111} & -0.0844\\
Gardnerella vaginalis & 0.6939 & \textbf{-2.9834} & -4.2448 & 0.4555 & \textbf{-0.9249} & -0.4500\\
Leptotrichia amnionii & 0.5094 & -2.7212 & -3.0925 & -0.2628 & -0.4553 & -0.3020\\
Bacteroides & 0.6935 & \textbf{-3.2685} & -5.5691 & -0.9732 & \textbf{0.1607} & 0.0274\\
Clostridiales & 0.4412 & -1.1113 & -4.1706 & 0.6240 & -0.8599 & 0.3289\\
Streptococcus anginosus & 1.1914 & \textbf{-2.7476} & -3.2483 & -0.4074 & \textbf{-0.3116} & -0.3012\\
Staphylococcus & -1.2574 & -1.8653 & -2.9380 & -0.4051 & -0.2712 & -0.5260\\
Bacteria & -0.8519 & -0.0419 & -5.5588 & -0.2858 & -0.2568 & 0.2760\\
Peptoniphilus lacrimalis & 0.0722 & -2.1754 & -4.4349 & -0.0621 & -0.5729 & 0.1260\\
Corynebacterium accolens & 0.2755 & \textbf{-3.5430} & -6.4496 & -0.1933 & \textbf{-0.1842} & 0.1541\\
Parvimonas micra & 1.7007 & -1.9734 & -4.0355 & -0.5192 & -0.1024 & 0.5413\\
Porphyromonas & -0.4227 & -2.2421 & -5.6416 & 0.4346 & -1.0118 & -0.3632\\
Prevotella bivia & 0.6292 & \textbf{-2.7653} & -4.0522 & 0.5408 & \textbf{-0.7516} & -0.4583\\
Bifidobacteriaceae & 1.2630 & -2.0531 & -4.9472 & 0.2806 & -0.1624 & -0.1074\\
Peptoniphilus asaccharolyticus & -0.1290 & \textbf{-3.6205} & -6.3497 & -0.6669 & \textbf{-0.2086} & -0.0705\\
Peptoniphilus harei & 0.3274 & \textbf{-2.4012} & -6.4962 & -0.4734 & \textbf{-0.2088} & 0.2916\\
Actinomyces & 1.7491 & \textbf{-2.8264} & -6.0772 & 0.2213 & \textbf{-0.5420} & -0.0433\\
Sneathia & 0.7347 & -4.7077 & -4.9721 & -0.7067 & -0.3644 & -0.7778\\
Gemella & 0.6244 & -3.1960 & -4.3473 & 0.0634 & -0.3158 & 0.1018\\
Coriobacteriaceae & 0.7464 & \textbf{-3.6256} & -5.1390 & -0.6120 & \textbf{-0.2043} & -0.2116\\
Veillonellaceae & 1.3996 & \textbf{-3.8063} & -6.2298 & -0.2779 & \textbf{0.1731} & -0.2848\\
Eggerthella & 3.9614 & \textbf{-3.8869} & -4.4172 & 0.2629 & \textbf{-1.0546} & -1.0317\\
Lachnospiraceae & 0.4803 & -3.3161 & -4.0073 & -1.3028 & -0.0050 & -0.7322\\
Bacteroidales & 0.8925 & \textbf{-2.6188} & -5.7584 & -0.5879 & \textbf{0.1703} & -1.1204\\
Peptostreptococcus & -1.5022 & -2.0752 & -3.6386 & -0.7837 & -0.3178 & -0.3577\\
Aerococcus christensenii & 0.8137 & \textbf{-1.7496} & -4.0424 & 0.6430 & \textbf{-0.6755} & -0.7022\\
Dialister sp type 3 & -0.4974 & \textbf{-2.8009} & -5.9637 & -0.5940 & \textbf{-0.4529} & -0.5076\\
Mobiluncus curtisii &  0.6626 & -4.2598 & -3.8239 & -0.8479 & -0.1577 & -0.7071\\
Lactobacillales & -0.0490 & \textbf{-3.7421} & -6.4767 & -0.6701 & \textbf{-0.1307} & 0.4199\\
Actinomycetaceae & -0.2150 & 0.0529 & -4.8075 & 0.3671 & -0.5665 & -0.6179\\
Anaerococcus tetradius & -1.8136 & -0.3787 & -4.3929 & 0.0235 & -0.5861 & -0.2021\\
Prevotella genogroup 3 & -3.1242 & 1.4811 & -4.4904 & -0.1353 & 0.7275 & 0.4336\\
Firmicutes & 0.2795 & -1.6765 & -5.7859 & -1.7511 & 1.1085 & -0.4444\\
Atopobium vaginae & 2.6754 & \textbf{-2.3296} & -1.8858 & 1.0732 & \textbf{-1.7974} & -0.7179\\
Streptococcus & 1.4210 & -2.3253 & -4.1598 & -0.6664 & -0.1194 & -0.6794\\
Aerococcus &  0.6551 & \textbf{-1.6029} & -5.4446 & 0.7116 & \textbf{-1.3571} & -3.0284\\
BVAB1 & -2.0914 & 2.6161 & -3.9562 & -0.3836 & -0.2357 & 0.1702\\
Ureaplasma & 0.6215 & -1.5126 & -5.5155 & -1.0830 & 0.0987 & -0.2251\\
\multicolumn{1}{l}{\textbf{More present in pregnant}}&&&&&&\\
Lactobacillus jensenii & 1.1874 & 2.6978 & -2.9872 & 0.4672 & 0.1064 & -0.6276\\
Lactobacillus crispatus & 1.2249 & \textbf{2.4304} & -2.1262 & 0.0366 & \textbf{1.2101} & 1.0443\\
Lactobacillus vaginalis & -1.2876 & 3.1935 & -5.5519 & 0.3050 & -0.2523 & -1.1646\\
Lactobacillus gasseri & 0.1602 & 0.3210 & -2.8409 & -0.8749 & 0.3214 & -0.1971\\
Lactobacillus & 2.8199 & 0.5166 & -2.7068 & -0.3749 & -0.5104 & -1.0172\\
Lactobacillus iners & 3.0629 & -0.2200 & -0.4422 & -0.2367 & 0.4068 & -0.8884\\
Prevotella genogroup 1 & -1.4204 & 3.3757 & -2.3026 & -1.4471 & -0.7525 & -2.6299\\
Megasphaera sp type 1 & 0.4711 & 1.2356 & -3.7876 & 0.3543 & -0.5830 & -0.2166\\
Sneathia sanguinegens & -0.3651 & \textbf{2.9115} & -3.0625 & -0.1183 & \textbf{-1.3089} & -1.6596\\
Proteobacteria & -0.3355 & 0.8134 & -6.0708 & 0.2224 & -0.2306 & 0.0222\\
\bottomrule
\end{tabular}

\begin{tablenotes}%
\scriptsize
\item Note: Bold coefficients represent a statistically significant variable for the corresponding taxa according to LRT ($\alpha=0.05$).
\item[$^{1}$] Variable scaled to $[0,1]$.
\end{tablenotes}
\end{threeparttable}
\end{table}

\begin{table}[htbp]
\footnotesize
\centering
\caption{ML estimates calculated by the SAEM algorithm for the parameters of Model 2 adjusted on the vaginal microbiome data}\label{aptab9}
\hspace*{-1cm}
\begin{threeparttable}
\begin{tabular}{rcccccccc}
\hline
&\multicolumn{4}{@{}c@{}}{\textbf{Logistic part}}&\multicolumn{4}{@{}c@{}}{\textbf{Beta part}}\\
\cmidrule{2-5}\cmidrule{6-9}
\multicolumn{1}{c}{\textbf{Taxa}} & \textbf{Time} & \textbf{Pregnancy} & \textbf{Age}$^{1}$ & \textbf{Interaction} & \textbf{Time} & \textbf{Pregnancy} & \textbf{Age}$^{1}$ & \textbf{Interaction}\\
\midrule
\multicolumn{1}{l}{\textbf{More present in non-pregnant}}&&&&&&&&\\
Actinomycetales & 0.3948 & -2.2550 & 1.2636 & -4.5273 & 0.2574 & -0.2449 & 0.4723 & -0.1987\\
Peptoniphilus & 0.3700 & 0.2810 & 0.9554 & \textbf{-5.1428} & 0.4979 & -0.0130 & 0.7428 & \textbf{-0.8741}\\
Finegoldia magna & 0.9772 & -0.3225 & -0.6225 & \textbf{-4.4261} & -0.1683 & -0.4308 & -0.0207 & \textbf{0.0193}\\
Anaerococcus & 0.8629 & -0.6215 & 1.8259 & -4.6912 & 0.0549 & 0.0580 & 0.4514 & -0.5976\\
Clostridiales Family XI Incertae Sedis & 0.7207 & -10.1685 & 1.4287 & -5.5151 & -0.1724 & -0.0642 & 0.5863 & 0.0701\\
Prevotella & 0.4429 & -1.3821 & 1.6417 & -4.8826 & 0.1590 & -0.1239 & 0.1221 & -0.8141\\
Anaerococcus vaginalis & 0.7951 & -4.0813 & -0.0944 & -5.4910 & -0.1770 & -0.1281 & 0.4443 & -0.1729\\
Prevotella genogroup 2 & 0.9482 & -0.5006 & 2.2294 & -4.4230 & 0.9075 & -0.3012 & 0.3405 & -1.2531\\
Dialister & 2.4420 & -2.6952 & 0.3451 & -5.7329 & 0.6666 & 0.2014 & 0.1664 & -0.9042\\
Atopobium & 3.3311 & -3.1559 & -2.0620 & \textbf{-5.4275} & -0.1449 & 0.0914 & 0.2228 & \textbf{-0.7851}\\
Gardnerella vaginalis & 0.6400 & -2.7894 & -0.8232 & -4.3367 & 0.5297 & -0.7076 & -0.2064 & -0.4432\\
Leptotrichia amnionii & 1.4334 & -0.6080 & -0.8310 & -3.2125 & -0.2064 & -0.0218 & -0.0096 & -0.5715\\
Bacteroides & 0.7602 & -5.7682 & 2.1606 & -5.7675 & -0.8673 & -7.3757 & 0.2836 & 9.7135\\
Clostridiales & 0.6560 & -0.9829 & 1.4696 & -4.3330 & \textbf{0.9404} & 1.0325 & 0.4875 & \textbf{-2.7669}\\
Streptococcus anginosus & 2.0122 & -0.3166 & 0.9938 & -3.3596 & -0.3247 & -0.3774 & -0.2174 & 0.1301\\
Staphylococcus & -1.5362 & -3.6995 & -0.6303 & -2.9711 & -0.4023 & -0.5308 & -0.4955 & 0.3844\\
Bacteria & -0.6773 & 0.2644 & 2.3964 & -5.5534 & -0.4084 & -0.5027 & 0.2986 & 0.4826\\
Peptoniphilus lacrimalis & -0.0278 & -2.0074 & 3.6833 & -4.6556 & -0.0100 & 1.4279 & 0.4354 & -2.5014\\
Corynebacterium accolens & 0.5597 & -0.7272 & 0.1489 & -6.5660 & -0.1154 & -0.5092 & 0.2607 & 0.6831\\
Parvimonas micra & 1.4761 & -2.9152 & 2.2397 & -4.3283 & -0.5811 & 0.0751 & 1.1713 & 0.0643\\
Porphyromonas & -0.1872 & 1.5996 & 2.2892 & -5.8198 & 0.4243 & -0.6574 & -0.1062 & -0.2790\\
Prevotella bivia & 1.6976 & 0.5426 & -1.3359 & \textbf{-4.1394} & 0.6937 & -0.3713 & -0.3830 & \textbf{-0.7431}\\
Bifidobacteriaceae & 1.7009 & -1.9078 & -0.6436 & -5.0737 & 0.2531 & -0.1764 & 0.1672 & 0.0999\\
Peptoniphilus asaccharolyticus & -0.0138 & -5.6375 & 1.3374 & -6.6523 & -0.5590 & -0.1947 & 0.3948 & 0.0481\\
Peptoniphilus harei & 0.7404 & 1.5115 & 1.1492 & -6.6702 & -0.3795 & -0.0607 & 0.5226 & -0.0205\\
Actinomyces & 1.3715 & \textbf{-7.1737} & 0.7501 & -6.2155 & 0.3464 & \textbf{-0.4793} & 0.1216 & -0.0661\\
Sneathia & 1.1231 & 0.1323 & 0.1920 & -5.1472 & -0.6467 & -0.5812 & -0.4221 & 0.4894\\
Gemella & 1.4290 & 0.5889 & -0.2813 & \textbf{-4.4960} & 0.3385 & 0.2498 & 0.1827 & \textbf{-1.1261}\\
Coriobacteriaceae & 1.0181 & -0.9975 & 1.5135 & -5.2714 & -0.5296 & -0.0980 & -0.0432 & 0.0134\\
Veillonellaceae & 2.0141 & 3.2373 & 1.9231 & \textbf{-6.2525} & -0.3010 & 4.9707 & -0.2761 & \textbf{-10.8236}\\
Eggerthella & 7.0001 & 1.3307 & 0.1952 & \textbf{-4.6386} & 0.7614 & 0.8781 & -0.8357 & \textbf{-3.1145}\\
Lachnospiraceae & 0.4188 & -6.9168 & 1.3680 & -4.1897 & -1.1530 & -0.0616 & -0.5525 & 0.0708\\
Bacteroidales & 0.7965 & -4.1245 & 2.7556 & -6.2071 & -0.3769 & 3.4525 & -0.4391 & -4.0362\\
Peptostreptococcus & -0.7720 & 0.7950 & 0.5703 & -3.7720 & -0.6830 & -0.4789 & -0.2520 & 0.3792\\
Aerococcus christensenii & 1.1763 & -1.3857 & -1.3277 & -4.2673 & 0.7233 & -0.5886 & -0.3450 & 0.0167\\
Dialister sp type 3 & -0.2289 & -1.6974 & 0.2424 & -6.1968 & -0.4251 & -0.2131 & -0.2086 & -0.2637\\
Mobiluncus curtisii & 0.3987 & -4.1474 & 1.6940 & -4.1145 & -1.0627 & 0.1107 & -0.1854 & 0.2001\\
Lactobacillales & -0.0543 & -4.3852 & 0.8537 & -6.4998 & -0.5817 & -0.1442 & 0.4201 & 0.0153\\
Actinomycetaceae & 0.2268 & 1.2868 & 2.9645 & -5.0411 & 0.3207 & -1.4314 & -0.2794 & 1.5222\\
Anaerococcus tetradius & -1.2706 & 1.1320 & 1.5860 & -4.6129 & 0.1162 & -0.0854 & 0.0456 & -0.7397\\
Prevotella genogroup 3 & -1.7386 & \textbf{5.3548} & 3.9245 & -4.4041 & -0.1681 & \textbf{1.4938} & 0.3868 & -1.8901\\
Firmicutes & 0.1821 & -3.8383 & 3.1218 & -5.9893 & -1.6163 & -3.2600 & -0.1226 & 5.5245\\
Atopobium vaginae & 3.3931 & -1.8104 & -0.2133 & -2.2539 & 1.5484 & -0.1086 & -0.2015 & -2.3932\\
Streptococcus & 2.2390 & -0.7838 & -2.3284 & -4.1217 & -1.0025 & -1.1272 & -0.6405 & 1.8209\\
Aerococcus & 1.0920 & \textbf{-0.1265} & 0.3653 & \textbf{-6.4798} & 0.4889 & \textbf{-1.1422} & -0.9858 & \textbf{1.0428}\\
BVAB1 & -2.4305 & 1.5174 & 0.0303 & -4.0814 & -0.2639 & 0.2136 & 0.2991 & -0.5858\\
Ureaplasma & 0.4596 & -1.5496 & -2.9635 & -5.5847 & -1.0195 & -0.0987 & -0.0569 & 0.2484\\
\multicolumn{1}{l}{\textbf{More present in pregnant}}&&&&&&&&\\
Lactobacillus jensenii & -0.4242 & \textbf{1.1176} & -1.0204 & \textbf{-3.1727} & -0.7723 & \textbf{-1.3386} & -0.1073 & \textbf{3.3769}\\
Lactobacillus crispatus & 1.5798 & \textbf{2.9938} & 1.7026 & -1.9845 & 0.0106 & \textbf{0.7945} & 0.8268 & 0.4186\\
Lactobacillus vaginalis & -3.5236 & -2.9340 & -5.3811 & \textbf{-6.1216} & -1.3030 & -0.7328 & 0.3448 & \textbf{3.0030}\\
Lactobacillus gasseri & -1.2666 & -1.9058 & -1.5392 & \textbf{-2.8785} & -0.9620 & 0.0504 & -0.0161 & \textbf{0.2846}\\
Lactobacillus & 2.6612 & \textbf{1.1124} & -2.7797 & -2.7404 & -0.6025 & \textbf{-1.0576} & -0.8576 & 1.0133\\
Lactobacillus iners & 2.9543 & -0.1484 & -2.8471 & -0.4604 & -0.2748 & 0.3779 & -0.8028 & 0.0509\\
Prevotella genogroup 1 & 1.4486 & \textbf{5.5025} & 1.2280 & \textbf{-2.9843} & -0.7541 & \textbf{0.5118} & -1.8807 & \textbf{-1.8349}\\
Megasphaera sp type 1 & 3.2792 & \textbf{3.7197} & 0.4440 & \textbf{-3.9302} & 0.4912 & \textbf{-0.6149} & -0.1020 & \textbf{0.0530}\\
Sneathia sanguinegens & 0.9750 & \textbf{4.0012} & 3.2933 & -3.1000 & -0.0251 & \textbf{-1.1921} & -1.6335 & -0.2152\\
Proteobacteria & -1.3540 & -0.9926 & 0.9526 & \textbf{-6.1615} & 0.2327 & -0.6051 & 0.1561 & \textbf{0.5710}\\
\bottomrule
\end{tabular}

\begin{tablenotes}%
\scriptsize
\item Note: Bold coefficients represent a statistically significant variable for the corresponding taxa according to LRT ($\alpha=0.05$).
\item[$^{1}$] Variable scaled to $[0,1]$.
\end{tablenotes}
\end{threeparttable}
\end{table}

\pagebreak

\section{Additional figures and tables}\label{secap4}

\subsection{Inflammatory bowel disorder pediatric study}\label{secap31}

\begin{figure}[htbp]%
\centering
\includegraphics[width=0.8\textwidth]{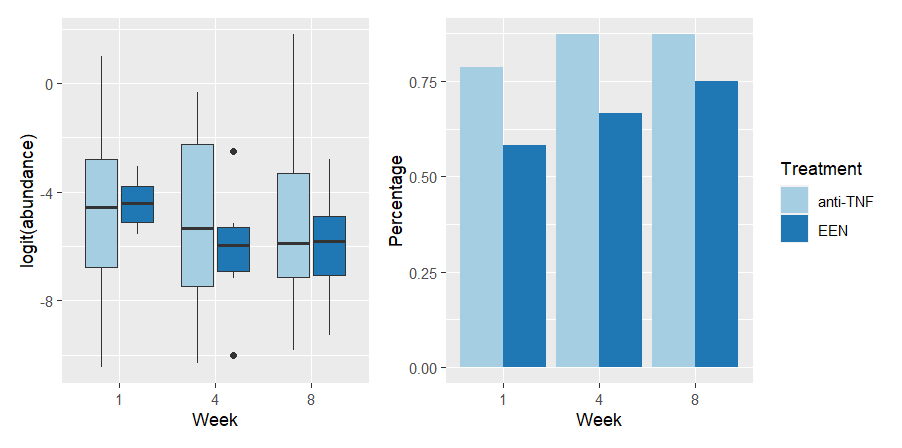}
\caption{Logit of the non-zero abundance (left) and percentage of samples with presence (right) for \textit{Escherichia} in each treatment group (anti-TNF and EEN) across observation week.}\label{apfig2}
\end{figure}

\begin{figure}[hbp]%
\centering
\includegraphics[width=0.9\textwidth]{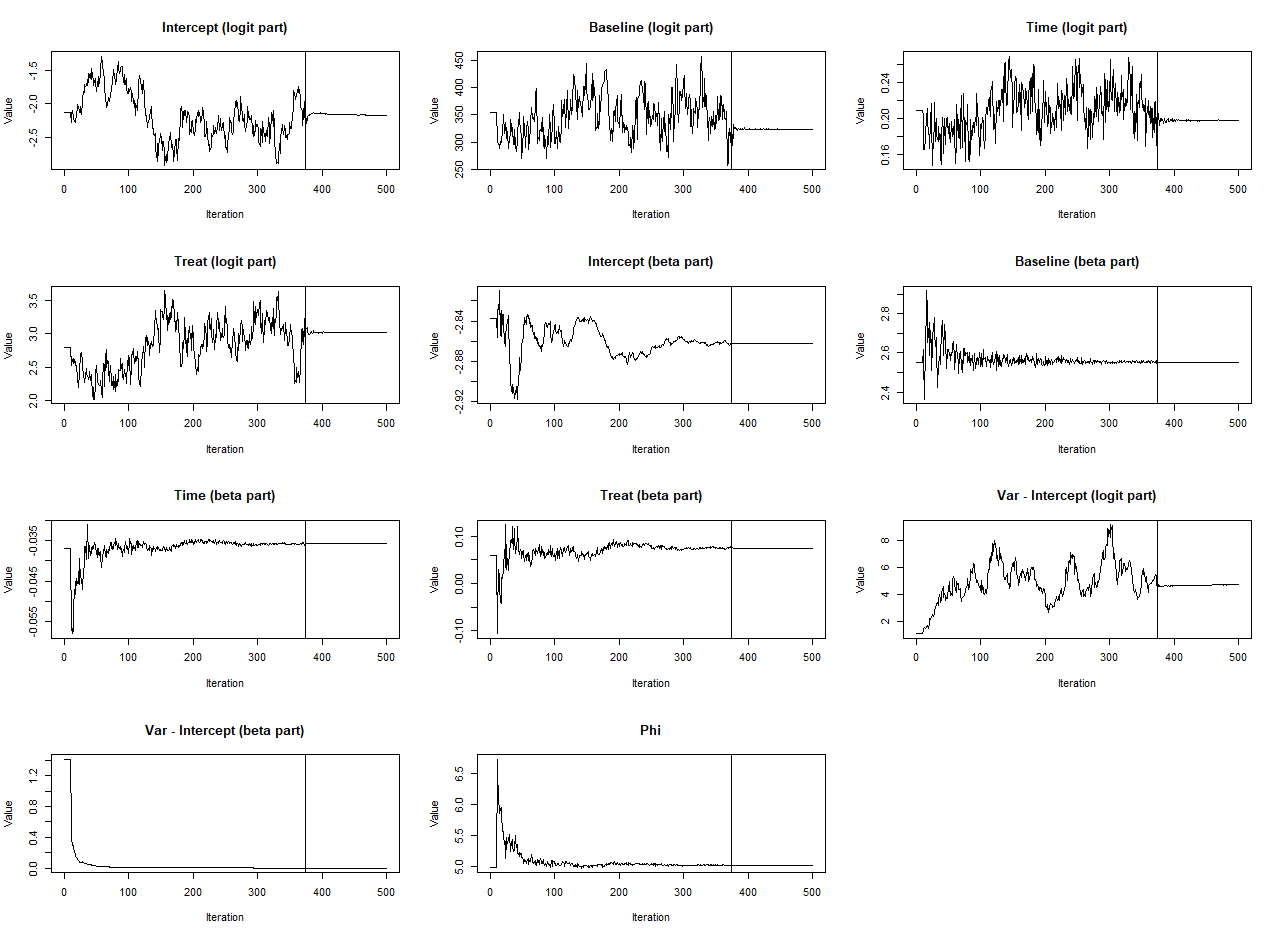}
\caption{Convergence of the ML estimates of the parameters of the ZIBR model for the \textit{Escherichia} genus calculated by the SAEM algorithm. The SAEM routine was implemented with 5 Markov chains and 500 iterations.}\label{apfig3}
\end{figure}

Table \ref{aptab5} shows the adjusted p-values for the variables considered in the model for each bacterial taxon, while Table \ref{aptab6} summarizes the estimators obtained by the ZIBR model for \textit{Escherichia}, clarifying in which part of the model the effect of the treatment is perceived. Furthermore, Figure \ref{apfig2} shows the convergence behavior of the SAEM algorithm for the ZIBR model in the mentioned taxon, evidencing one of its great advantages, which is not needing a large number of iterations when entering the phase of descent to the final values. Figure \ref{apfig3} supports the results of Table \ref{aptab6}, where the evolution over time of both the presence or absence of \textit{Escherichia} and its average abundance in individuals is seen, showing a differentiated behavior between treatments and controls in the presence but not in the abundance.

\begin{table}[htbp]
\centering
\caption{P-values obtained by the Likelihood Ratio Test based on the SAEM algorithm for the bacterial taxa of the IBD patients data. The p-values were corrected using the Benjamini-Hochberg process to decrease the false discovery rate.\vspace*{2mm}}\label{aptab5}
\begin{tabular}{lccc}

\hline
\textbf{Species} & \textbf{Baseline} & \textbf{Time} & \textbf{Treat}\\
\midrule
Bacteroides & 0.0000 & 0.1172 & 0.2133\\
Ruminococcus & 0.0008 & 0.1203 & 0.0033\\
Faecalibacterium & 0.0000 & 0.2645 & 0.0009\\
Bifidobacterium & 0.0000 & 0.1621 & 0.0008\\
Escherichia & 0.0000 & 0.0293 & 0.0308\\
Clostridium & 0.0003 & 0.2905 & 0.0934\\
Dialister & 0.0008 & 0.2227 & 0.0045\\
Eubacterium & 0.0049 & 0.0106 & 0.0114\\
Roseburia & 0.0000 & 0.1587 & 0.0708\\
Streptococcus & 0.0170 & 0.1706 & 0.0000\\
Dorea & 0.0000 & 0.3773 & 0.0605\\
Parabacteroides & 0.0000 & 0.0989 & 0.2028\\
Lactobacillus & 0.0000 & 0.2208 & 0.0020\\
Veillonella & 0.0000 & 0.7333 & 0.0053\\
Haemophilus & 0.0000 & 0.1961 & 0.0000\\
Alistipes & 0.0000 & 0.4010 & 0.0000\\
Collinsella & 0.0000 & 0.1228 & 0.0078\\
Coprobacillus & 0.0000 & 0.5255 & 0.3298\\
\bottomrule
\end{tabular}
\end{table}

\begin{table}[htbp]
\centering
\caption{Estimated effects with the SAEM algorithm of the variables in the ZIBR model for \textit{Escherichia}.\vspace*{2mm}}\label{aptab6}
\begin{threeparttable}
\begin{tabular*}{0.8\textwidth}{@{\extracolsep\fill}lcc}
\hline%
\textbf{Variable}$^{1}$&\textbf{Beta part} & \textbf{Logistic part} \\
\midrule
Baseline&2.5521***&323.66***\\
&(0.3656)&(89.4662)\\
Time&-0.0356&0.1979***\\
&(0.0261)&(0.0412)\\
Treat&0.0746&3.0186***\\
&(0.1576)&(0.8815)\\
\bottomrule
\end{tabular*}
\vspace*{6mm}
\begin{tablenotes}
\footnotesize
\item Note: Standard errors of the respective coefficients in parentheses. Symbol * (**,***) represents significance at 10\% (5\%, 1\%) level.
\item[$^{1}$] Statistical significance is calculated with the Wald test.\vspace*{6pt}
\end{tablenotes}
\end{threeparttable}
\end{table}

\end{appendices}

\pagebreak